\documentclass[twocolumn, twocolappendix, times]{aastex63}
\usepackage{graphicx}
\usepackage{amsmath}

\submitjournal{Research in Astronomy and Astrophysics}
\shortauthors{Kuznetsov et al.}
\shorttitle{Observations of AT Mic with AstroSat}

\begin{document}
\title{X-Ray and Ultraviolet Flares on AT Microscopii Observed by AstroSat}

\correspondingauthor{Alexey A. Kuznetsov}
\email{a\_kuzn@iszf.irk.ru}

\author[0000-0001-8644-8372]{Alexey A. Kuznetsov}
\affiliation{Institute of Solar-Terrestrial Physics, Irkutsk, 664033, Russia}
\affiliation{Department of Geography, Irkutsk State University, Irkutsk, 664033, Russia}

\author{Ruslan R. Karakotov}
\affiliation{Institute of Solar-Terrestrial Physics, Irkutsk, 664033, Russia}

\author[0000-0001-5803-0649]{Kalugodu Chandrashekhar}
\affiliation{Rosseland Centre for Solar Physics, University of Oslo, PO Box 1029 Blindern, 0315 Oslo, Norway}
\affiliation{Institute of Theoretical Astrophysics, University of Oslo, PO Box 1029 Blindern, 0315 Oslo, Norway}

\author[0000-0003-4653-6823]{Dipankar Banerjee}
\affiliation{Aryabhatta Research Institute of Observational Sciences (ARIES), Manora Peak, Nainital-263 002, India}
\affiliation{Indian Institute of Astrophysics, Sarjapur Main Road, 2nd Block, Bangalore, Karnataka 560034, India}
\affiliation{Center of Excellence in Space Science, IISER Kolkata, Kolkata 741246, India}

\begin{abstract}
We present observations of the active M-dwarf binary AT Mic (dM4.5e+dM4.5e) obtained with the orbital observatory AstroSat. During 20 ks of observations, in the far ultraviolet ($130-180$ nm) and soft X-ray ($0.3-7$ keV) spectral ranges, we detected both quiescent emission and at least five flares on different components of the binary. The X-ray flares were typically longer than and delayed (by $5-6$ min) with respect to their ultraviolet counterparts, in agreement with the Neupert effect. Using X-ray spectral fits, we have estimated the parameters of the emitting plasma. The results indicate the presence of a hot multi-thermal corona with the average temperatures in the range of $\sim 7-15$ MK and the emission measure of $\sim (2.9-4.5)\times 10^{52}$ $\textrm{cm}^{-3}$; both the temperature and the emission measure increased during the flares. The estimated abundance of heavy elements in the corona of AT Mic is considerably lower than at the Sun ($\sim 0.18-0.34$ of the solar photospheric value); the coronal abundance increased during the flares due to chromospheric evaporation. The detected flares had the energies of $\sim 10^{31}-10^{32}$ erg; the energy-duration relations indicate the presence of magnetic fields stronger than in typical solar flares.
\end{abstract}


\section{Introduction}
Many dwarf stars, including the Sun, demonstrate signatures of magnetic activity, such as starspots, hot coronae, and flares \citep{Gershberg2005}. Cool red dwarfs of dMe spectral type are usually much more active than the Sun, which is manifested in stronger magnetic fields, hotter and denser coronae with a persistent presence of nonthermal electrons, and more frequent and powerful flares \citep{Haisch1991, Gudel2002, Gudel2004, Reiners2012}; nevertheless, the physical mechanisms responsible for the coronal heating and flares are believed to be qualitatively similar to the solar ones. 

Observations in the soft X-ray range provide an opportunity to study the thermal plasma in the stellar coronae and determine its parameters, such as density, temperature, emission measure, and chemical composition -- both in the quiescent state and during flares \citep{Gudel2004}. On the other hand, optical and UV emissions of flares are produced in deeper layers of stellar atmospheres (the chromosphere and transition region), in response to heating of these layers by nonthermal electron beams \citep{Benz2010}; the optical continuum is responsible for the most part of the radiated flare energy \citep{Kretzschmar2011}. Simultaneous observations in different spectral ranges (e.g., soft X-rays and UV) are of special interest, because they allow one to study the processes at different layers of a stellar atmosphere and, in particular, investigate correlations between thermal and nonthermal processes. While the Sun is continuously observed by many instruments providing a broad spectral coverage (from radio to $\gamma$-rays), multiwavelength observations of flares on other stars are much less common. Thus expanding the dataset of such observations (both for the same or for different targets) is highly important, because it improves the reliability of the derived physical models, as well as allows one to study long-term variations of the stellar active phenomena, including activity cycles.

In this work, we investigate the well-known active M-dwarf binary AT Mic using observations with the orbital observatory AstroSat; manifestations of stellar activity were detected in X-ray and UV spectral ranges. We analyze the characteristics of the coronal plasma in the quiescent and flaring states; we also investigate the relationship between the X-ray and UV flares. We compare the obtained results with earlier observations of AT Mic and other red dwarfs, as well as with observations and models of solar flares.

\section{Observations}
\subsection{Target}
AT Mic is a visual binary consisting of two almost identical red dwarfs of spectral type dM4.5e \citep{Joy1974}, at a distance of 9.86 pc \citep{Gaia2020}. The binary is sufficiently wide, with a semi-major axis of $\sim 31$ au and an orbital period of about 209 years \citep{Malkov2012}, so that interaction between the components is negligible; i.e., in terms of magnetic activity, they can be considered as two separate single stars. AT Mic is believed to be a member of the $\beta$ Pictoris association, with an age of $\sim 25$ Myr \citep{Messina2017}. The rotation periods of the components A and B are of 1.19 and 0.78 days, respectively \citep{Messina2016, Messina2017}; the bolometric luminosities of the components A and B have been estimated as $1.3\times 10^{32}$ and $1.2\times 10^{32}$ erg $\textrm{s}^{-1}$ (0.034 and 0.031 $L_{\sun}$), respectively \citep{Messina2017}. Both components are known to be magnetically active \citep[e.g.,][]{Gershberg1999}.

Due to its proximity and high activity level, AT Mic has been a target of many observations in different spectral ranges. Powerful optical flares with energies of up to $\sim 4\times 10^{33}$ erg were reported by \citet{Kunkel1970, Nelson1986, GarciaAlvarez2002}, etc. \citet{Pallavicini1990, Raassen2003, Robrade2005}, etc. observed AT Mic in the soft X-ray range and detected quiescent emission of hot coronal plasma with a luminosity of $\sim 2\times 10^{29}$ erg $\textrm{s}^{-1}$, as well as a number of flares with energies of up to $\sim 1.3\times 10^{33}$ erg. \citet{Linsky1982, Elgaroy1988, MonsignoriFossi1995}, etc. detected quiescent chromospheric emission and flares in the UV range, while \citet{MitraKraev2005} investigated correlations between X-ray and UV flares. The durations of the detected flares varied from less than one minute up to about one hour, with the above mentioned energies corresponding to the longest and most powerful flares.\\[12pt]

\subsection{Observations with AstroSat}
We observed AT Mic with the instruments onboard the AstroSat satellite \citep{Singh2014, Agrawal2017a}, on 2018 October 3-4. The primary instrument was the Soft X-ray Telescope (SXT) -- an X-ray telescope with focusing mirrors that provides X-ray imaging and spectroscopy in the nominal energy range of $0.3-8$ keV with the energy channel width of 0.01 keV \citep{Singh2016, Singh2017}. The SXT operated in the photon counting mode; the total exposure time was 20 ks. A half of that time was also covered by high-resolution imaging observations with the Ultra-Violet Imaging Telescope (UVIT) \citep{Tandon2017a, Tandon2017b} in the far ultraviolet channel (FUV, $130-180$ nm), in the photon counting mode; the F148W filter with the $125-175$ nm bandpass was used. The target was detected by both of these instruments, thus providing simultaneous observations in the two spectral ranges for $\sim 10$ ks.

Simultaneously, AT Mic was observed with two other X-ray detectors: the Large Area X-ray Proportional Counter (LAXPC) \citep{Yadav2016, Agrawal2017b} in the $3-100$ keV energy range, and the Cadmium Zinc Telluride Imager (CZTI) \citep{Bhalerao2017} in the $25-150$ keV energy range. However, no reliable signal of stellar origin was detected by these instruments, due to either insufficient sensitivity (for the CZTI, given that the X-ray flux decreases rapidly with energy) or a high and rapidly varying instrumental background (for the LAXPC); we do not analyze these observations here.

Multiwavelength observations (including the soft X-ray and UV spectral ranges) of AT Mic and other active red dwarfs have been performed earlier with the XMM-Newton observatory \citep[e.g.,][]{MitraKraev2005, Tsang2012, Perdelwitz2018}. In comparison with the XMM-Newton European Photon Imaging Camera (EPIC) \citep{Struder2001, Turner2001}, the AstroSat SXT has a lower sensitivity to the soft X-rays, which requires longer accumulation times or time bins to obtain a reliable signal from the same source. On the other hand, in comparison with the XMM-Newton Optical Monitor (OM) \citep{Mason2001}, the AstroSat UVIT has considerably better sensitivity and angular resolution in the far UV range, which allows one to study the morphology and light curves of UV sources with much finer details.

\begin{figure}
\centerline{\includegraphics{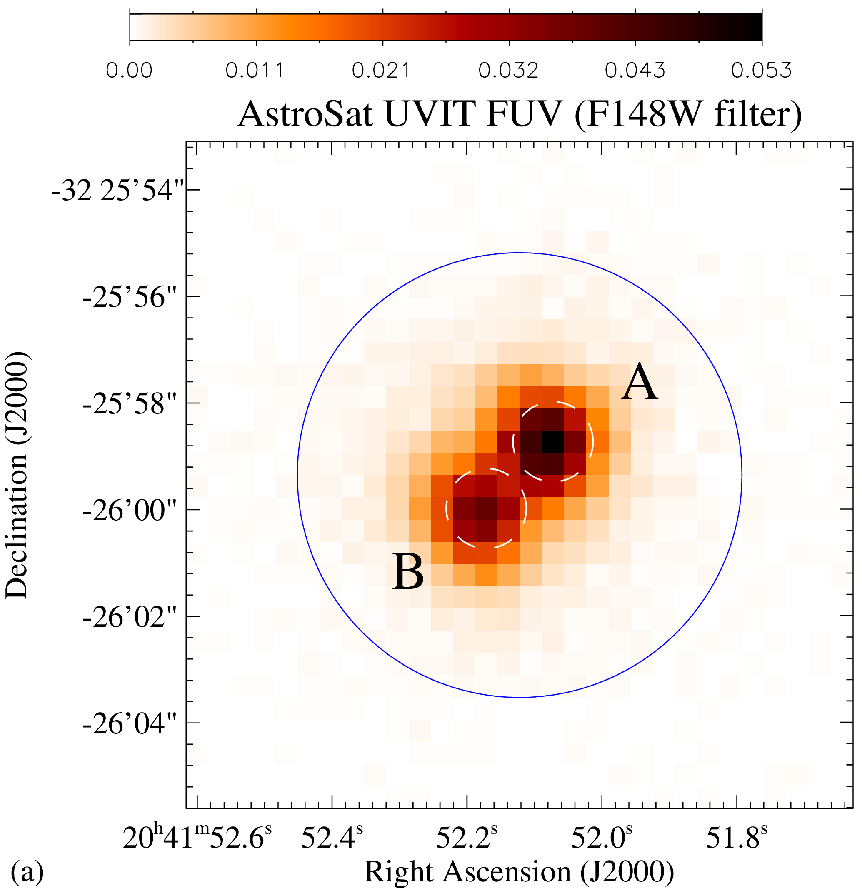}}
\centerline{\includegraphics{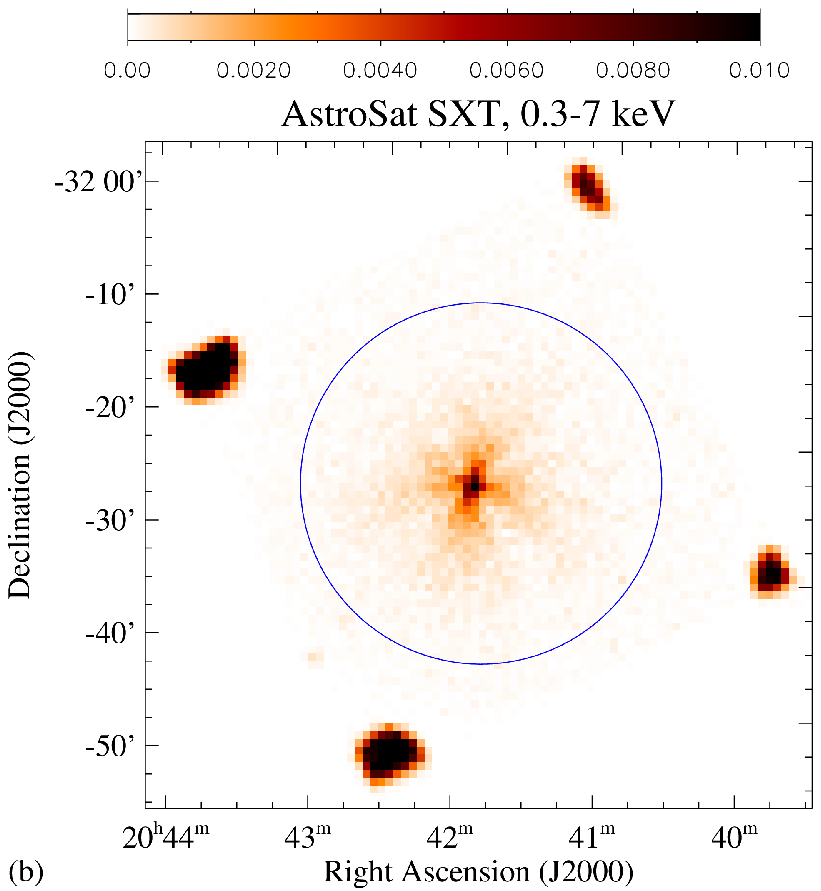}}
\caption{UV (a) and X-ray (b) images of AT Mic from the AstroSat UVIT FUV and SXT observations on 2018 October 3-4, for the entire observation intervals. The color bars are in units of counts $\textrm{s}^{-1}$ $\textrm{pixel}^{-1}$. The blue or white circles show the extraction regions for light curves and spectra.}
\label{FigImages}
\end{figure}

\section{Results}
\subsection{Images}
Figure \ref{FigImages} demonstrates the images of AT Mic obtained with the AstroSat UVIT FUV and SXT, integrated over the entire duration of the observations. Figure \ref{FigImages}a shows a fragment of the UVIT field of view; the binary components are partially resolved with the component A (the northern one) being slightly brighter. Figure \ref{FigImages}b shows the full SXT field of view (photons in the energy range of $0.3-7$ keV were selected, because at higher energies the signal might be contaminated by instrumental noise); the bright spots at the corners of the detector matrix are the internal calibration sources. The X-ray photons are scattered over a relatively wide area, so that the binary components are not resolved; nevertheless, the X-ray centroid position coincides with the AT Mic position with an accuracy of $\sim 1.5'$, and there are no other candidate X-ray sources in this region of the sky.

The circles in Figure \ref{FigImages} show the regions used to extract the UV and X-ray light curves and X-ray spectra. For the UV emission, we mainly consider the total flux from both components of the binary, within a radius of $4.2''$. Since the components are resolved only partially, we cannot separate the fluxes from them completely; the partial fluxes from the individual binary components (from the regions shown by white dashed circles with a diameter of $1.5''$, corresponding to the overall FWHM of the UVIT FUV) are analyzed only qualitatively -- to identify the component where a flare occurred. For the total X-ray flux, we use the extraction region with a radius of $16'$, which contains more than 96.5\% of the source photons (according to the point spread function presented in the AstroSat Handbook\footnote{\url{http://www.iucaa.in/~astrosat/AstroSat_handbook.pdf}}).

\begin{figure*}
\centerline{\includegraphics{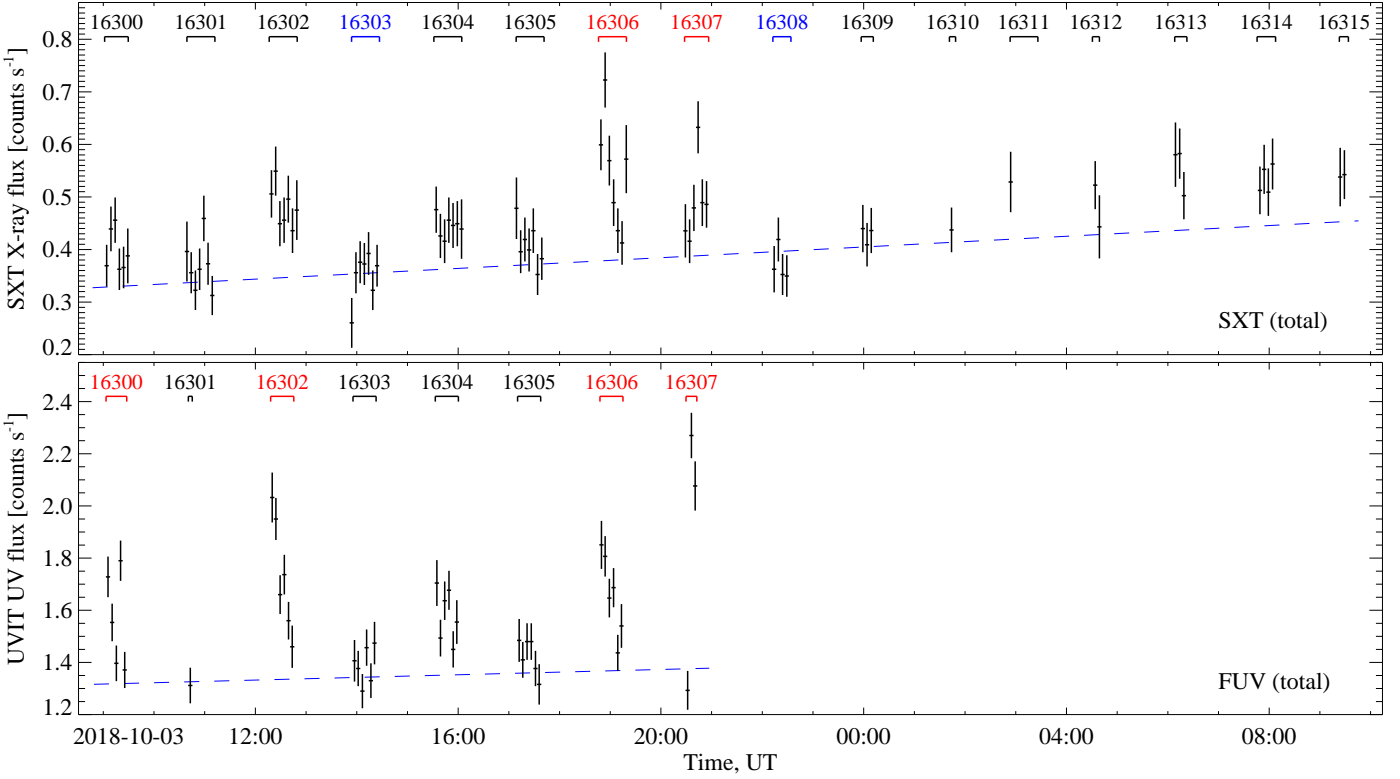}}
\caption{X-ray (top) and UV (bottom) light curves of AT Mic from the AstroSat SXT and UVIT FUV observations on 2018 October 3-4, with 300 s time bins. The error bars correspond to the $1\sigma$ level. The blue dashed lines represent the estimated quiescent (non-flaring) background fluxes. The AstroSat orbit numbers are shown at the tops of the panels, with the supposed flaring or (for X-rays) quiescent-only time intervals indicated by red or blue colors, respectively.}
\label{FigLClong}
\end{figure*}

\subsection{Light Curves}
\subsubsection{Light curves: overview}
Figure \ref{FigLClong} demonstrates the total (i.e., combining the fluxes from both components of the binary) X-ray and UV light curves of AT Mic, with the time bin size of 300 s. Due to periodic occultations of the target by the Earth, the light curves are not continuous but consist of a number of shorter time intervals (orbits); due to stricter visibility constraints, the UVIT FUV observing intervals are shorter than those for the SXT, and the UVIT observations ceased after the first eight orbits. For the X-ray light curve, photons in the energy range of $0.3-7$ keV were selected and the instrumental background provided by the SXT team\footnote{\url{https://www.tifr.res.in/~astrosat_sxt/dataanalysis.html}\label{dataanalysis}} was subtracted; for the UV observations, the instrumental background is negligible.

\begin{deluxetable*}{ccccccccccc}
\renewcommand{\tabcolsep}{2.6pt}
\tablewidth{0pt}
\tablecaption{Parameters of the flares detected on AT Mic: sources (i.e., components of the binary where the flares occurred), decay times in the X-ray ($\tau_{\mathrm{X}}$) and UV ($\tau_{\mathrm{UV}}$) ranges, delays of the X-ray flares with respect to the corresponding UV flares ($\Delta t_{\mathrm{X}-\mathrm{UV}}$), peak luminosities in the X-ray range ($L_{\mathrm{X}}^{\max}$) and in the optical continuum ($L_{\mathrm{cont}}^{\max}$), emitted energies in the X-ray range ($E_{\mathrm{X}}$) and in the optical continuum ($E_{\mathrm{cont}}$), and estimated \protect\citep[following the scaling laws by][]{Namekata2017} sizes of the flaring regions ($L$) and magnetic field strengths in these regions ($B$).\label{Flares}}
\tablehead{\colhead{Flare} & \colhead{Source} & \colhead{$\tau_{\mathrm{X}}$,} & \colhead{$\tau_{\mathrm{UV}}$,} & \colhead{$\Delta t_{\mathrm{X}-\mathrm{UV}}$,} & \colhead{$L_{\mathrm{X}}^{\max}$,} & \colhead{$L_{\mathrm{cont}}^{\max}$,} & \colhead{$E_{\mathrm{X}}$,} & \colhead{$E_{\mathrm{cont}}$,} & \colhead{$L$,} & \colhead{$B$,}\\[-6pt]
\colhead{} & \colhead{} & \colhead{min} & \colhead{min} & \colhead{min} & \colhead{$10^{29}$ erg $\textrm{s}^{-1}$} & \colhead{$10^{29}$ erg $\textrm{s}^{-1}$} & \colhead{$10^{31}$ erg} & \colhead{$10^{31}$ erg} & \colhead{$10^9$ cm} & \colhead{G}}
\startdata
F1 & B & $ 7.88{\pm}  7.57$ & $ 3.67{\pm}  1.44$ & $ 4.87{\pm}  1.72$ & $  0.46_{-0.19}^{+0.21}$ & $  0.92_{-0.62}^{+1.19}$ & $ 2.19_{-0.68}^{+0.77}$ & $ 3.03_{-1.94}^{+3.50}$ & $ 4.46_{-1.48}^{+1.48}$ & $101.1_{-33.6}^{+58.0}$\\
F2 & B & $\cdots$ & $ 2.78{\pm}  1.15$ & $ 5.21{\pm}  1.72$ & $\cdots$ & $  1.19_{-0.77}^{+1.43}$ & $\cdots$ & $ 2.20_{-1.41}^{+2.56}$ & $ 3.75_{-1.28}^{+1.27}$ & $112.0_{-37.9}^{+67.7}$\\
F3 & B & $ 7.45{\pm}  5.90$ & $ 6.54{\pm}  2.68$ & $ 0.57{\pm}  1.72$ & $  0.64_{-0.21}^{+0.23}$ & $  1.35_{-0.87}^{+1.58}$ & $ 4.14_{-0.93}^{+0.99}$ & $ 6.71_{-4.10}^{+6.89}$ & $ 6.59_{-2.17}^{+2.12}$ & $ 83.8_{-27.4}^{+48.7}$\\
F4 & A, B & $10.62{\pm}  3.97$ & $\cdots$ & $\gtrsim (5.61{\pm}  1.72)$ & $  1.05_{-0.26}^{+0.31}$ & $  1.27_{-0.84}^{+1.60}$ & $ 8.71_{-1.40}^{+1.68}$ & $> 5.27_{-3.27}^{+5.64}$ & $\cdots$ & $\cdots$\\
F5 & A & $ 9.31{\pm}  7.00$ & $ 4.30{\pm}  0.85$ & $ 5.07{\pm}  1.72$ & $  0.73_{-0.22}^{+0.26}$ & $  2.69_{-1.65}^{+2.81}$ & $ 3.77_{-0.88}^{+1.05}$ & $\gtrsim 6.81_{-4.12}^{+6.82}$ & $\gtrsim 5.59_{-1.34}^{+1.31}$ & $\gtrsim 108.0_{-27.5}^{+33.6}$\\
\enddata
\tablecomments{The peak luminosities ($L_{\mathrm{X}}^{\max}$ and $L_{\mathrm{cont}}^{\max}$) and emitted energies ($E_{\mathrm{X}}$ and $E_{\mathrm{cont}}$) of the flares were estimated using Equations (\protect\ref{FlareEnergy}--\protect\ref{L_cont}), as described in Section \protect\ref{EnergyEstimations}.}
\end{deluxetable*}

Identifying the flaring and non-flaring (quiescent) time intervals in our observations is not straightforward, because of the above mentioned gaps in the light curves, and also because of a high activity level: AT Mic spends in a flaring state $\sim 25\%$ of time, according to \citet{Messina2016}, or even more than $50\%$ of time, according to \citet{Kunkel1970}. We determined the quiescent background levels by iterative fitting the light curves with linear functions and removing the data points that exceeded these fits by more than $1\sigma$; this procedure effectively provides a linear fit to the lower envelope of a light curve. The X-ray quiescent background tends to increase with time -- probably, reflecting the stellar rotation and/or evolution of active regions; the UV quiescent background also demonstrates a weak increase with time.

We consider the time intervals where the UV and/or X-ray fluxes exceeded the respective quiescent background levels by more than $3\sigma$ as candidate flares. We have identified five such flaring events scattered over four AstroSat orbits (namely, orbits \#16300, \#16302, \#16306, and \#16307; see the next Section for details). In addition, we have selected in the X-ray light curve two time intervals with the lowest flux (namely, orbits \#16303 and \#16308), which we consider as a ``quiescent state'', although these intervals could still contain weaker unresolved flares.\\[18pt]

\begin{figure*}
\centerline{\includegraphics{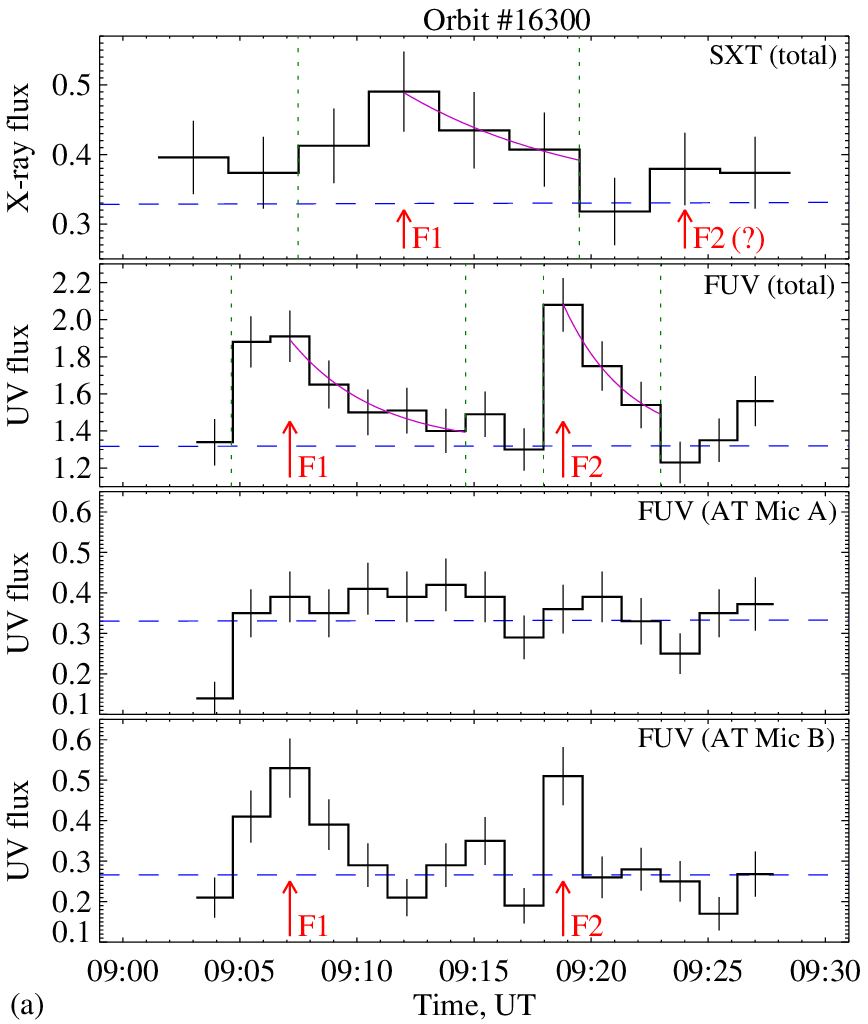}
\includegraphics{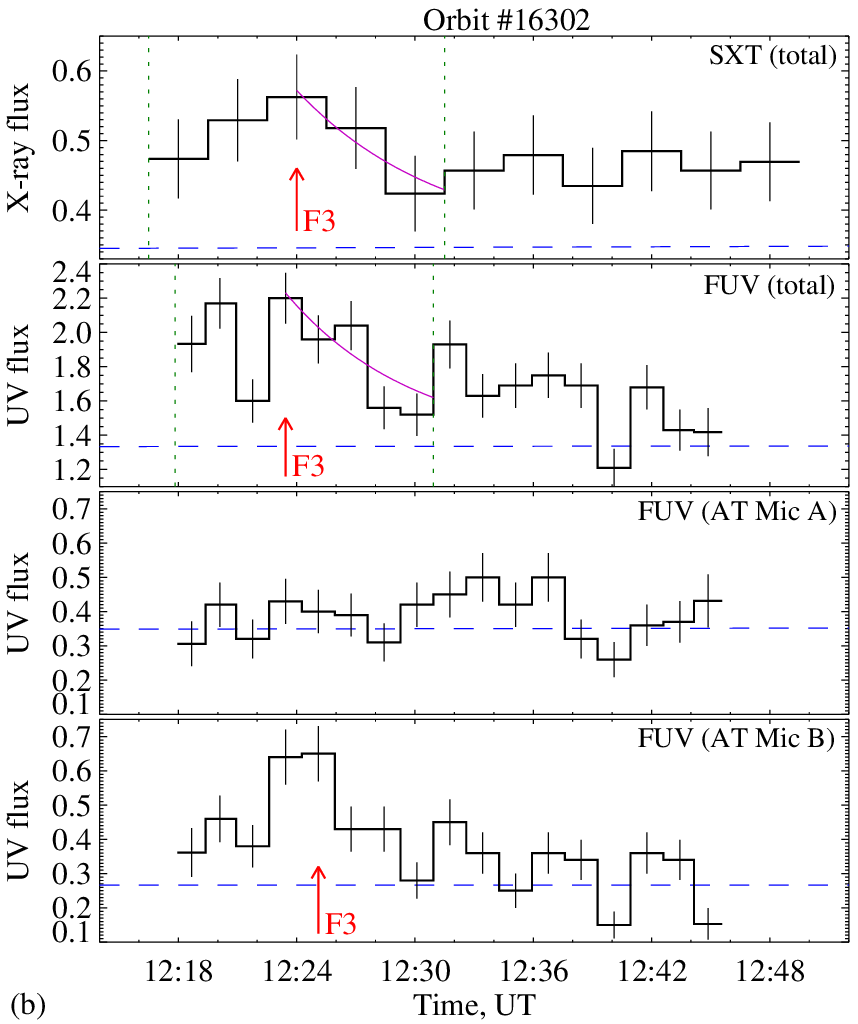}}
\centerline{\includegraphics{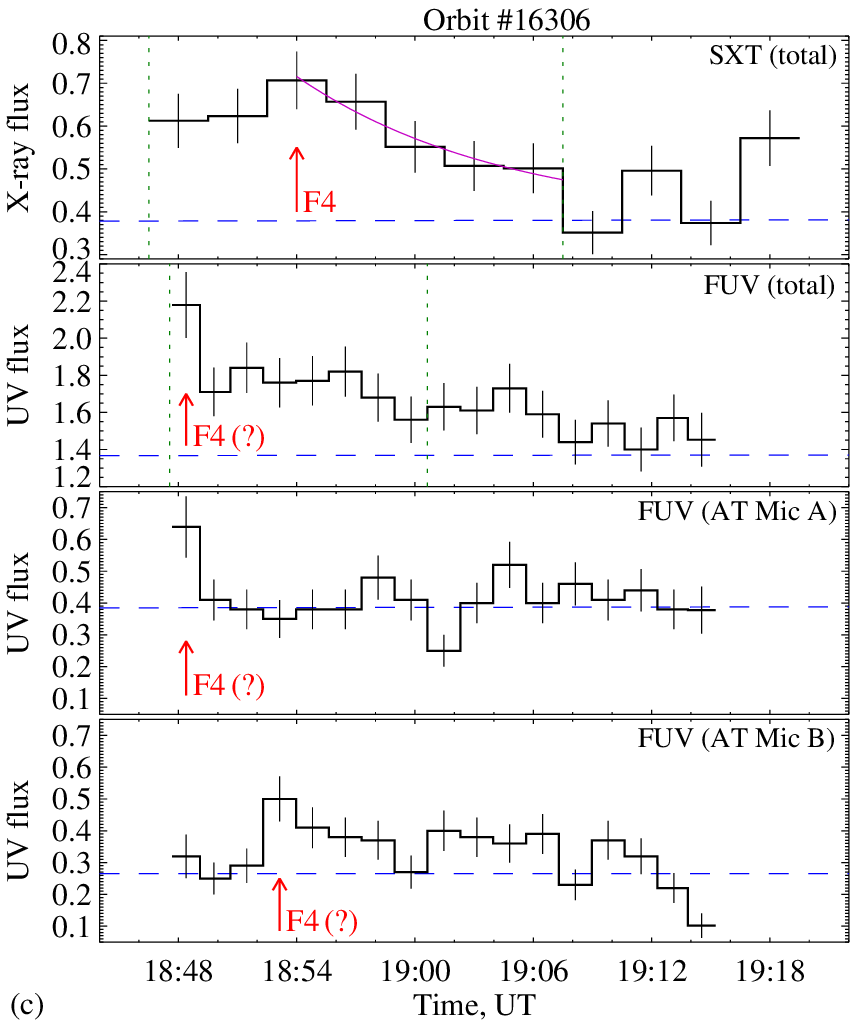}
\includegraphics{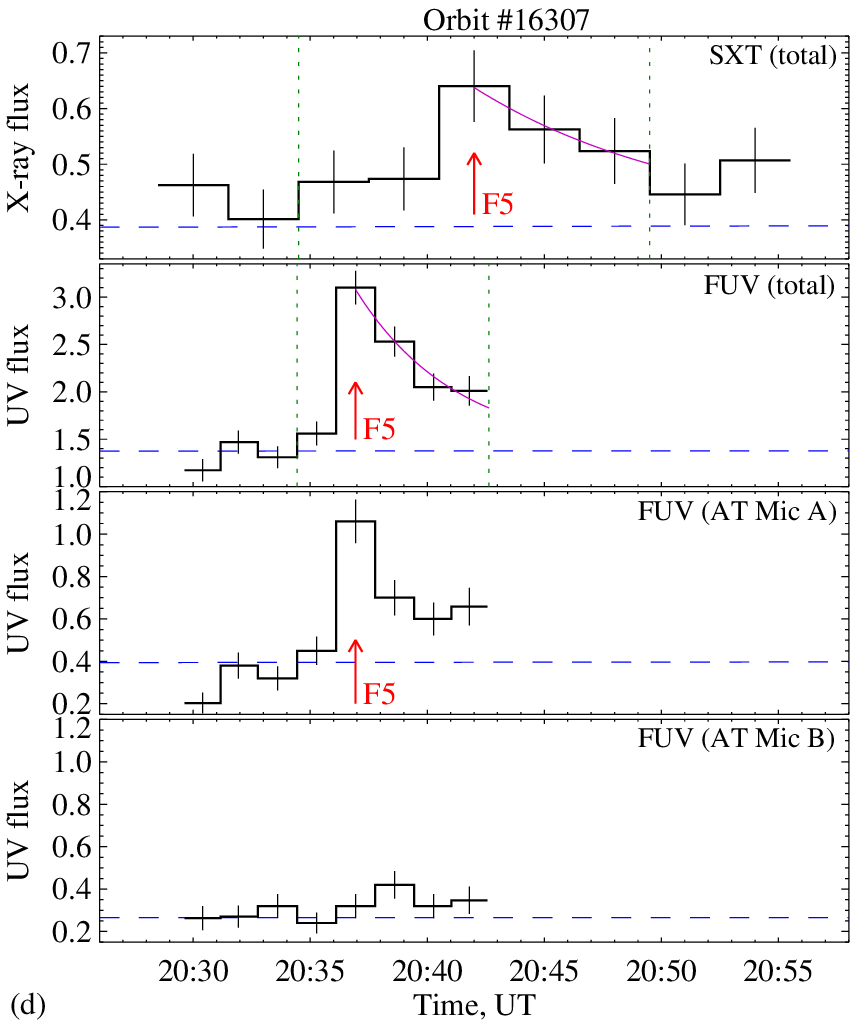}}
\caption{X-ray and UV light curves for the selected time intervals (cf. Figure \protect\ref{FigLClong}) where flaring activity was detected. For the UV emission, both the total flux from the AT Mic system and the fluxes corresponding to its individual components are shown. The fluxes are in counts $\textrm{s}^{-1}$; the time bins are 180 s and 100 s for the X-ray and UV light curves, respectively. The error bars correspond to the $1\sigma$ level. The flare IDs used in the text are shown next to the red arrows. The vertical dotted lines show the flare start/end times used in the analysis, while the continuous magenta curves represent the exponential fits (\protect\ref{decay}) to the light curves at the flare decay phases.}
\label{FigLCshort}
\end{figure*}

\subsubsection{Light curves of flares}
We show in Figure \ref{FigLCshort} enlarged fragments of the light curves for the time intervals containing flares. Together with the total (unresolved) X-ray and UV light curves of AT Mic, we show the UV light curves corresponding to the individual components of the binary. The time bins are reduced (in comparison to Figure \ref{FigLClong}) to 180 s for the X-rays and 100 s for the UV emission. The basic characteristics of the detected flares are summarized in Table \ref{Flares}. In particular, the $e$-folding decay times of the flares $\tau_{\mathrm{X}}$ and $\tau_{\mathrm{UV}}$ (which we use to characterize the flare durations) are estimated by least-squares fitting the flare decay phases in the total light curves shown in Figure \ref{FigLCshort} with exponential functions in the form
\begin{equation}\label{decay}
I(t)=I_{\mathrm{bg}}(t)+A\exp\left(-\frac{t-t_{\mathrm{peak}}}{\tau}\right),\qquad
t\ge t_{\mathrm{peak}},
\end{equation}
where $I(t)$ is the flux in the considered spectral band, $I_{\mathrm{bg}}(t)$ is the estimated quiescent background flux, and $t_{\mathrm{peak}}$ is the flare peak time. The delays between the X-ray and UV flares $\Delta t_{\mathrm{X}-\mathrm{UV}}$ are defined as delays between the respective flare peaks; the uncertainty in estimating the delays $\sigma_{\mathrm{X-UV}}$ is estimated as $\sigma_{\mathrm{X-UV}}^2=(\Delta t_{\mathrm{X}}/2)^2+(\Delta t_{\mathrm{UV}}/2)^2$, where $\Delta t_{\mathrm{X}}$ and $\Delta t_{\mathrm{UV}}$ are the time bins of the X-ray and UV light curves, respectively.

In the total UV light curve in Figure \ref{FigLCshort}a, one can see two well-defined flares, peaked at around 09:07 and 09:19 UT, both with ``classical'' flare profiles consisting of a sharp rise and a slower exponential decay. Comparison with the light curves of the individual binary components indicates that both flares occurred on the component B. The flares are relatively short -- with durations of just a few minutes. In the X-ray light curve, one can see a flare (again, with a ``classical'' profile) peaked at around 09:12 UT, which likely corresponds to the UV flare F1; however, the X-ray flare is longer than (with the characteristic decay time $\tau_{\mathrm{X}}\simeq 2\tau_{\mathrm{UV}}$) and delayed by $\sim 5$ min with respect to its UV counterpart. Later, at around 09:24 UT, there is a barely noticeable X-ray peak -- possibly, a counterpart of the UV flare F2. We cannot determine reliably the decay time of this supposed X-ray flare; however, like in the previous event, the X-ray peak is delayed with respect to the UV flare peak by $\sim 5$ min.

In the total light curves in Figure \ref{FigLCshort}b, one can see a flare (F3) that peaked at around 12:24 UT, nearly simultaneously in the X-ray and UV ranges; the UV flare occurred on the component B. In contrast to the previous two flares, the UV flare F3 has a more complicated profile -- likely, a result of overlapping of several partially resolved flaring events. The X-ray flare F3 has a nearly triangular shape with comparable rise and decay times. The delay between the X-ray and UV peaks is almost absent, and the flare durations in both spectral ranges are similar ($\tau_{\mathrm{X}}\simeq \tau_{\mathrm{UV}}$).

The time interval shown in Figure \ref{FigLCshort}c contains the largest and longest X-ray flare (F4) among the detected ones, with a ``classical'' flare profile, peaked at around 18:54 UT. Interpretation of the UV light curves is less straightforward: at the beginning of the time interval, one can see a UV peak -- likely, a tail of a powerful flare occurred on the component A; if this UV flare indeed corresponds to the X-ray flare F4, the delay between the X-ray and UV flares is at least of $5-6$ min. Later, at around 18:53 UT, there is a weaker UV flare on the component B, which, however, is not pronounced in the total UV light curve. Because of a complicated time profile, we cannot estimate reliably the duration of the UV flare. The profile of the X-ray flare is also most likely a result of overlapping of two flares occurred on different components of the binary.

Finally, the time interval shown in Figure \ref{FigLCshort}d contains another well-defined X-ray flare (F5) and its UV counterpart, both with ``classical'' flare profiles, peaked at around 20:42 and 20:37 UT, respectively; the UV flare occurred on the component A. The UV flare was observed only partially; nevertheless, we can estimate its characteristic decay time. The X-ray flare is longer than (with $\tau_{\mathrm{X}}\simeq 2\tau_{\mathrm{UV}}$) and delayed by $\sim 5$ min with respect to the corresponding UV flare.

\subsection{X-Ray Spectral Analysis}\label{analysis}
We analyzed the SXT X-ray spectra in the energy range of $0.3-7$ keV using the OSPEX package \citep{OSPEX} and the SXT instrumental background spectrum and response files provided by the SXT team\textsuperscript{\ref{dataanalysis}}. Model X-ray spectra were fitted to the observations using the Markov Chain Monte Carlo (MCMC) approach; the MCMC sampling implementation by \citet{Anfinogentov2021} was used. Among several spectral models (see the examples and discussion in Appendix \ref{MCMC}), the best agreement with the observations was achieved using the model \textsf{multi\_therm\_gauss} that describes optically thin bremsstrahlung radiation from multi-thermal plasma with a Gaussian dependence of the differential emission measure (DEM) on the logarithm of the temperature ($T$):
\begin{equation}\label{DEMmodel}
\mathrm{DEM}(T)=\mathrm{DEM}_0\exp\left[-\frac{(\log T-\log T_0)^2}{2\sigma_T^2}\right],
\end{equation}
with $\mathrm{DEM}_0$, $T_0$, and $\sigma_T$ being the parameters of the Gaussian. The corresponding total emission measure (EM) and average plasma temperature $(\left<T\right>)$ are given by the formulae:
\begin{equation}
\mathrm{EM}=\left(\sqrt{2\pi}\ln 10\right)\mathrm{DEM}_0T_0\sigma_T\exp\left(\frac{1}{2}\sigma_T^2\ln^210\right),
\end{equation}
\begin{equation}\label{TavgDEM}
\left<T\right>=T_0\exp\left(\frac{3}{2}\sigma_T^2\ln^210\right).
\end{equation}
The spectral model depended also on the abundance of heavy elements $Z$, relative to the solar abundance; the ratios between the elements were fixed to the solar values.

\begin{figure}
\centerline{\includegraphics{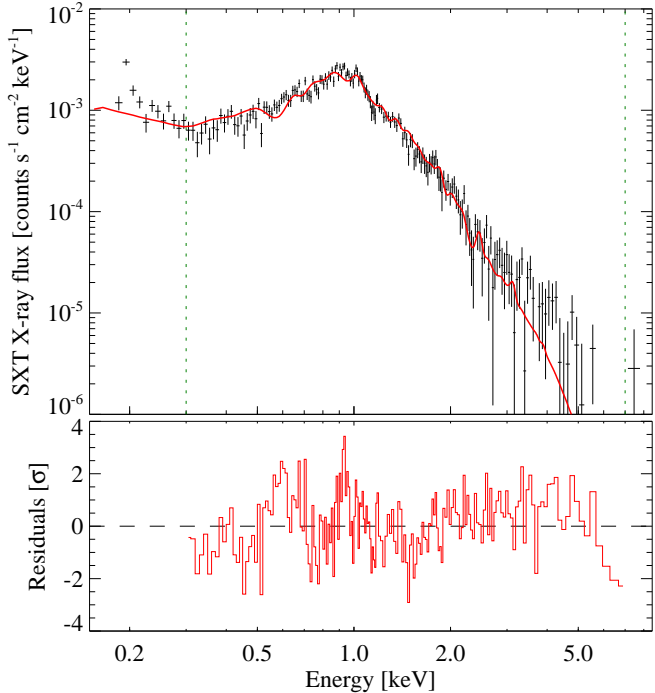}}
\caption{Top: X-ray spectrum of AT Mic from the AstroSat SXT observations on 2018 October 3-4, averaged over the entire interval of observations. At higher energies, the original spectrum was rebinned over $2-65$ channels ($0.02-0.65$ keV). The error bars correspond to the $1\sigma$ level. The solid red line represents a model spectral fit with the multi-thermal plasma emission model given by Equation (\protect\ref{DEMmodel}). Bottom: normalized residuals of the model spectral fit, in units of the $1\sigma$ uncertainties.}
\label{FigSpectrum}
\end{figure}

To account for the interstellar absorption, we used the model by \citet{Morrison1983}. The absorption column density was fixed to the value of $N_{\mathrm{H}}=3\times 10^{19}$ $\textrm{cm}^{-2}$ that was estimated using the optical extinction of $A_V=0.017$ mag reported by \citet{Malkov2012} and the empirical relation of $N_{\mathrm{H}}=1.79\times 10^{21}A_V$ $\textrm{cm}^{-2}$ $\textrm{mag}^{-1}$ by \citet{Predehl1995}. The effect of the interstellar absorption was found to be minor, and considering the absorption column density as a free parameter has not significantly affected the obtained results.

\begin{deluxetable}{lcccc}
\renewcommand{\tabcolsep}{3.5pt}
\tablewidth{0pt}
\tablecaption{Parameters of the X-ray spectral fits for the selected time intervals: emission measures (EM), average temperatures ($\left<T\right>$), relative widths of the DEM distribution ($\sigma_T$), and abundances of heavy elements ($Z$).\label{SXTfits}}
\tablehead{\colhead{} & \colhead{EM,} & \colhead{$\left<T\right>$,} & \colhead{$\sigma_T$,} & \colhead{$Z$,}\\[-6pt]
\colhead{} & \colhead{$10^{52}$ $\textrm{cm}^{-3}$} & \colhead{keV} & \colhead{$\log(\mathrm{keV})$} & \colhead{$Z_{\sun}$}}
\startdata
Total & $3.54_{-0.11}^{+0.12}$ & $0.728_{-0.020}^{+0.028}$ & $0.168_{-0.016}^{+0.021}$ & $0.192_{-0.014}^{+0.021}$\\
Quiescent & $2.92_{-0.29}^{+0.29}$ & $0.637_{-0.033}^{+0.053}$ & $0.131_{-0.047}^{+0.050}$ & $0.176_{-0.031}^{+0.070}$\\
\hline
Flare F1 & $3.33_{-0.65}^{+0.60}$ & $0.680_{-0.054}^{+0.169}$ & $0.117_{-0.058}^{+0.122}$ & $0.185_{-0.047}^{+0.192}$\\
Flare F3 & $3.21_{-0.53}^{+0.45}$ & $1.028_{-0.100}^{+0.464}$ & $0.238_{-0.060}^{+0.100}$ & $0.338_{-0.062}^{+0.282}$\\
Flare F4 & $4.42_{-0.43}^{+0.66}$ & $1.199_{-0.159}^{+1.335}$ & $0.384_{-0.074}^{+0.300}$ & $0.273_{-0.056}^{+0.156}$\\
Flare F5 & $3.45_{-0.47}^{+0.61}$ & $0.842_{-0.078}^{+0.417}$ & $0.214_{-0.054}^{+0.167}$ & $0.313_{-0.074}^{+0.205}$\\
\enddata
\end{deluxetable}

We performed the spectral fitting of the SXT spectra in several time intervals, including all reliably detected X-ray flares (i.e., the flares F1 and F3-F5, with the used start/end times of the flaring intervals shown by vertical dotted lines in the X-ray light curves in Figure \ref{FigLCshort}) and the ``quiescent'' state (defined as a combination of two AstroSat orbits with the lowest X-ray flux, i.e. orbits \#16303 and \#16308), as well as for the entire (total) duration of the observations. An example of the SXT X-ray spectrum of AT Mic, together with the best-fit spectral model, is shown in Figure \ref{FigSpectrum} (we note that, for illustration purposes, the spectrum in Figure \ref{FigSpectrum} was rebinned to reduce the number of points at high energies, while the spectral analysis was performed using the original spectral resolution). The obtained best-fit parameters of the emitting plasma are summarized in Table \ref{SXTfits} and demonstrated in Figure \ref{FigSXTfits}.

\section{Discussion}
\subsection{Light Curves}
\subsubsection{Neupert effect}
As can be seen in Figure \ref{FigLCshort} and Table \ref{Flares}, light curves of flares F1 and F5 (and, possibly, flare F2) demonstrate the so-called Neupert effect \citep{Neupert1968}, when the soft X-ray flares are longer than and delayed with respect to the corresponding UV flares; the delays ($\sim 5$ min) are similar in all three flares. The major UV flare F4 (on AT Mic A) was observed only partially; nevertheless, the light curves indicate a delay of at least $5-6$ min between the X-ray and UV peaks, in agreement with the Neupert effect. In the ``standard'' flare model \citep[e.g.,][]{Benz2010}, this effect occurs because the optical or UV emissions represent a direct (and immediate) response of the chromosphere to heating by nonthermal electrons, while the soft X-rays are produced by heated plasma evaporated from the chromosphere to the corona, with the emission intensity proportional to the cumulative (time-integrated) nonthermal energy input.

On the other hand, the flare F3 demonstrates no significant delay between the X-ray and UV peaks, and the flare durations in both spectral ranges are similar, i.e., the Neupert effect is not pronounced. Probably, this is caused by a complicated (with several separate peaks, as indicated by the UV light curve) dynamics of energy release in this flare, when the chromospheric/coronal responses to separate acts of particle acceleration are mixed together.

The Neupert effect is often (but not always) observed in solar and stellar flares \citep[e.g.,][]{Benz2010}. \citet{MitraKraev2005}, using soft X-ray and UV observations with XMM-Newton, detected the Neupert effect in flares on several red dwarfs, including AT Mic, although the reported X-ray-to-UV flare delays on AT Mic ($\sim 17$ min) were longer than in our observations. Most likely, this difference is caused by the selection effect: due to a higher sensitivity and hence shorter time bins of the AstroSat UVIT (in comparison with the XMM-Newton OM), the flares detected in the UV range in this study were much shorter and weaker than those reported by \citet{MitraKraev2005}, and had respectively shorter delays between the X-ray and UV peaks.

\subsubsection{Activity levels of the binary components}
In total, in the UV range (where the binary components were partially resolved) we detected four flares on the component B and one or two (if we count here the partially observed flare F4) flares on the component A, i.e., the component B was considerably more active. A similar conclusion follows indirectly from the spatially resolved observations of AT Mic in the microwave range reported by \citet{Kundu1987}: in that work, flaring activity was detected on both components of the binary, with the southern component (AT Mic B) demonstrating a stronger variability and higher microwave fluxes, although individual flares could not be identified due to short duration of the observations and insufficient time resolution. If confirmed, a higher magnetic activity of AT Mic B is most likely related to the faster rotation of this component, which results in a more efficient dynamo action \citep[e.g.,][]{Brun2017}. On the other hand, we note that the number of detected flares in our study was too small to make a statistically significant conclusion.\\[18pt]

\begin{figure}
\centerline{\includegraphics{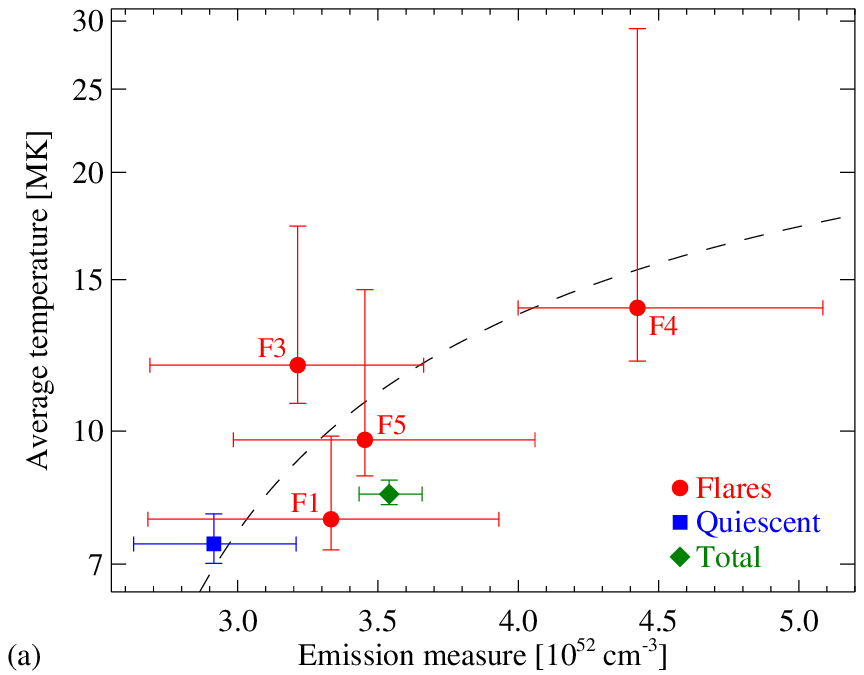}}
\centerline{\includegraphics{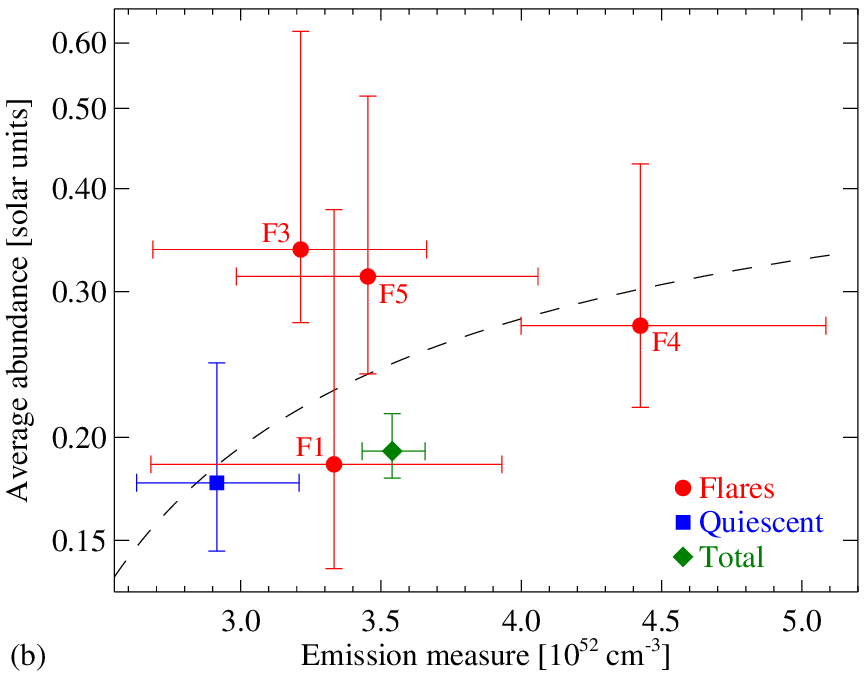}}
\caption{X-ray spectral fit parameters: average plasma temperatures (a) and abundances (b) vs. emission measures in the corona of AT Mic for different time intervals. The error bars correspond to the $1\sigma$ level. The dashed lines in panels (a) and (b) represent model fits given by Equations (\protect\ref{Tavg}) and (\protect\ref{Zavg}), respectively.}
\label{FigSXTfits}
\end{figure}

\subsection{X-Ray Spectral Fits}
\subsubsection{Coronal temperatures and emission measures}
Figure \ref{FigSXTfits}a shows the results of the X-ray spectral analysis -- the emission measures and average plasma temperatures for different time intervals (we note that these parameters are actually either summed or averaged over both components of the binary). In the supposedly quiescent state, the emission measure and average temperature are of $\sim 2.9\times 10^{52}$ $\textrm{cm}^{-3}$ and $\sim 7.4$ MK, respectively. During flares, both the emission measure and average temperature increase considerably, while the time-averaged parameters for the entire duration of the observations are intermediate between the quiescent and flaring values.

Since we analyze the total emission spectra and do not separate the flaring and non-flaring spectral components, the obtained plasma temperatures during flares actually represent weighted averages of the temperatures in the quiescent coronae and in the flaring regions, which can be expressed as
\begin{equation}\label{Tavg}
\left<T\right>=\frac{T_{\mathrm{q}}\mathrm{EM}_{\mathrm{q}}+T_{\mathrm{flare}}(\mathrm{EM}-\mathrm{EM}_{\mathrm{q}})}{\mathrm{EM}},
\end{equation}
where $T_{\mathrm{q}}$ and $\mathrm{EM}_{\mathrm{q}}$ are respectively the temperature and emission measure in the quiescent state, and $T_{\mathrm{flare}}$ is the plasma temperature in a flaring region. We have fitted Equation (\ref{Tavg}) to the observations assuming that the temperatures $T_{\mathrm{flare}}$ in all flaring regions are approximately the same (which is a very crude assumption) and excluding the ``total'' data point; the best agreement has been achieved for $T_{\mathrm{flare}}$ of about $31.8\pm 20.0$ MK. Similar temperatures ($\sim 7-8$ MK and $\sim 28-34$ MK, respectively) of the quiescent and flaring plasma components in the coronae of AT Mic were reported earlier in the papers of \citet{Raassen2003} and \citet{Robrade2005}, where these components were resolved spectrally.

\subsubsection{Coronal abundances}
Figure \ref{FigSXTfits}b shows the emission measures and coronal abundances of heavy elements for different time intervals. In the supposedly quiescent state, the coronal abundance takes its lowest value of about 0.18 of the solar photospheric abundance. The photospheric abundance for AT Mic is not exactly known; however, assuming it to be similar to that for other members of the $\beta$ Pictoris association \citep[$\sim 1.12$ of the solar one for $\beta$ Pic itself, according to][]{Gray2006}, we conclude that the corona of AT Mic is considerably depleted in heavy elements in comparison to the photosphere. 

As can be seen from Figure \ref{FigSXTfits}b, the coronal abundance tends to increase during flares. Most likely, this increase is caused by the chromospheric evaporation: during flares, the material from lower atmospheric layers is heated and expands upward into the coronal flaring loops, thus enriching the corona with heavy elements (we remind that in this study the measured abundances during flares actually represent weighted averages of the abundances in the quiescent coronae and in the flaring regions). The time-averaged coronal abundance ($\sim 0.19$ of the solar photospheric abundance) is intermediate between the quiescent and flaring values, and is consistent with the coronal abundances for AT Mic reported by \citet{Robrade2005}. An increase of the coronal abundance during flares has been observed earlier on M dwarfs \citep[e.g.,][]{Favata2000, Liefke2010}, RS CVn binaries \citep[e.g.][]{Pandey2012}, and other active stars \citep[e.g,][]{Gudel2001}.

Similarly to Equation (\ref{Tavg}), the average coronal abundance during flares (affected by the chromospheric evaporation) can be expressed in the form
\begin{equation}\label{Zavg}
\left<Z\right>=\frac{Z_{\mathrm{q}}\mathrm{EM}_{\mathrm{q}}+Z_{\mathrm{chromo}}(\mathrm{EM}-\mathrm{EM}_{\mathrm{q}})}{\mathrm{EM}},
\end{equation}
where $Z_{\mathrm{q}}$ and $Z_{\mathrm{chromo}}$ are the coronal abundance in the quiescent state and the chromospheric abundance, respectively. We have fitted Equation (\ref{Zavg}) to the observations (excluding the ``total'' data point); the best agreement has been achieved for the chromospheric abundance $Z_{\mathrm{chromo}}$ of about $0.53\pm 0.32$ of the solar photospheric abundance. We note that the obtained value of $Z_{\mathrm{chromo}}$ should be considered as an effective one, because the used model does not account for the dynamics of the chromospheric evaporation process as well as for the physical processes responsible for the underabundance of heavy elements in the corona.\\[18pt]

\subsection{Flare Energies}
\subsubsection{Energy estimations}\label{EnergyEstimations}
We estimate the total radiated energy of a stellar flare $E^{\mathrm{flare}}$ as
\begin{equation}\label{FlareEnergy}
E^{\mathrm{flare}}=\sum\limits_t L^{\mathrm{flare}}(t)\Delta t,
\end{equation}
where $L^{\mathrm{flare}}(t)$ is the time-dependent flare luminosity in the considered spectral band (see below) and $\Delta t$ is the width of the time bin. The summation was performed over the time intervals bounded by vertical dotted lines in the total light curves in Figure \ref{FigLCshort}. The flare luminosity in the X-ray range is given by \citep[see][]{MitraKraev2005, Kuznetsov2021}
\begin{equation}
L^{\mathrm{flare}}_{\mathrm{X}}(t)=\frac{\left<L_{\mathrm{X}}\right>}{\left<I_{\mathrm{X}}\right>}\left[I_{\mathrm{X}}(t)-I_{\mathrm{X}}^{\mathrm{bg}}(t)\right],
\end{equation}
where $I_{\mathrm{X}}(t)$ is the total X-ray light curve (counts $\textrm{s}^{-1}$), $I_{\mathrm{X}}^{\mathrm{bg}}(t)$ is the corresponding quiescent background X-ray flux, $\left<I_{\mathrm{X}}\right>$ is the average total X-ray flux in the selected time interval, and $\left<L_{\mathrm{X}}\right>$ is the average total X-ray luminosity in the selected time interval. The latter quantity is estimated using a spectral fit:
\begin{equation}
\left<L_{\mathrm{X}}\right>=2\pi d^2\int\limits_{E_{\min}}^{E_{\max}}F(E)E\,\mathrm{d}E,
\end{equation}
where $d$ is the distance to the target, $F(E)$ is the model X-ray spectral flux density, and the flux in the spectral range from $E_{\min}$ to $E_{\max}$ is considered ($0.3-7$ keV in this work); this formula represents the energy flux from an optically thin source into the upper hemisphere. The peak X-ray luminosities $L_{\mathrm{X}}^{\max}$ of the flares detected on AT Mic (computed using the spectral fits described in Section \ref{analysis}) are presented in Table \ref{Flares}.

The flare luminosity in the AstroSat FUV spectral band is given by
\begin{equation}\label{L_FUV}
L^{\mathrm{flare}}_{\mathrm{FUV}}(t)=\pi d^2G\left[I_{\mathrm{FUV}}(t)-I_{\mathrm{FUV}}^{\mathrm{bg}}(t)\right],
\end{equation}
where $I_{\mathrm{FUV}}(t)$ is the total UV light curve (counts $\textrm{s}^{-1}$), $I_{\mathrm{FUV}}^{\mathrm{bg}}(t)$ is the corresponding quiescent background UV flux, and $G$ is the count rate to flux conversion factor which equals $1.545\times 10^{-12}$ erg $\textrm{count}^{-1}$ $\textrm{cm}^{-2}$ for the UVIT FUV instrument with the F148W filter\footnote{\url{https://uvit.iiap.res.in/Instrument/Filters}}; this formula represents the energy flux from an optically thick flat source into the upper hemisphere. While the AstroSat SXT spectral range covers most of the thermal X-ray radiation from stellar flares, the AstroSat UVIT FUV spectral band contains only a small fraction of a flare radiation. Following \citet{Brasseur2019} and \citet{Fleming2022}, we assume that the UV radiation of a flare can be described as a blackbody radiation of hot flare ribbons, and estimate the bolometric flare luminosity $L_{\mathrm{cont}}^{\mathrm{flare}}$ by extrapolating the luminosity in the FUV spectral band (\ref{L_FUV}) to the entire optical continuum:
\begin{equation}\label{L_cont}
L_{\mathrm{cont}}^{\mathrm{flare}}=L_{\mathrm{FUV}}^{\mathrm{flare}}\frac{\int\limits_0^{\infty}B(\lambda, T_{\mathrm{eff}})\,\mathrm{d}\lambda}{\int\limits_{\lambda_1}^{\lambda_2}B(\lambda, T_{\mathrm{eff}})\,\mathrm{d}\lambda},
\end{equation}
where $B(\lambda, T)$ is the Planck function, $T_{\mathrm{eff}}$ is the effective temperature of the flare ribbons, and the wavelength range in the denominator corresponds to the bandpass of the AstroSat UVIT FUV F148W filter ($125-175$ nm). We use $T_{\mathrm{eff}}=10\,000$ K as the typical effective temperature of the UV-emitting flare ribbons \citep{Kowalski2013}, with possible variations in the range of $9\,000-12\,000$ K. The extrapolation (\ref{L_cont}) is very sensitive to the adopted temperature $T_{\mathrm{eff}}$, so that the uncertainty in this value is the main source of uncertainties in the estimated bolometric luminosities and radiated energies of stellar flares. The peak estimated bolometric luminosities $L_{\mathrm{cont}}^{\max}$ of the flares detected on AT Mic are presented in Table \ref{Flares}.

We note, however, that the bolometric luminosity estimation (\ref{L_cont}) is approximate, because, in addition to a blackbody component, the UV emission of flares can contain a significant contribution of line emission and/or Balmer continuum. As demonstrated by \citet{Kowalski2019}, in the near ultraviolet range ($\gtrsim 250$ nm) the blackbody model may underestimate the flare flux by a factor of about two; at shorter wavelengths (including the AstroSat UVIT FUV spectral band), the contribution of non-blackbody components has not been determined yet. Thus, a more accurate FUV-to-bolometric conversion would require spectroscopic observations.

The estimated radiated energies of the detected flares in the soft X-ray range and in the optical continuum are presented in Table \ref{Flares}. For the flare F4, only the late part was observed in the UV range; therefore, the radiated energy of this flare in the optical continuum was likely considerably (up to several times) higher than the value shown in Table \ref{Flares}. Also, because the decay phase of the UV flare F5 was observed only partially, the radiated energy of this flare in the optical continuum shown in Table \ref{Flares} is likely underestimated by $\sim 30$\% (assuming an exponential decay); this difference, however, is smaller than the uncertainties caused by other factors.

\subsubsection{Flare energy partition}
As follows from Table \ref{Flares}, the radiated energies of the detected flares were of $\sim 10^{31}-10^{32}$ erg, i.e., these flares were comparable to the strongest known solar flares \citep[e.g.][]{Shibata2002, Emslie2005}, but far less powerful than the strongest flares observed on AT Mic before \citep{Pallavicini1990, GarciaAlvarez2002}. For the three flares (F1, F3, and F5), where the radiated energies both in the X-ray range and in the optical continuum are known with a sufficient accuracy, we can compare the flare energy outputs in these spectral channels: the optical continuum dominated, with $E_{\mathrm{cont}}/E_{\mathrm{X}}\sim 1.4-1.8$, and was responsible for $\sim 60-65\%$ of the total radiated flare energy. This energy partition is consistent with that in solar flares, where the white-light continuum is responsible, on average, for $\sim 70$\% of the total radiated flare energies \citep{Kretzschmar2011}. Similar relations between the optical and X-ray flare emissions on other stars were reported, e.g., by \citet{Guarcello2019, Schmitt2019, Kuznetsov2021}.

Similarity of the flare energy partitions obtained in this and other works confirms that the above-described approach to estimate the optical continuum luminosity is justified, i.e. a) the choice of the effective temperature of flare ribbons to be $T_{\mathrm{eff}}=10\,000$ K is adequate, and b) contribution of non-blackbody components in the FUV spectral band is minor.

\subsection{Active Region Parameters}
To estimate the parameters of the active regions on AT Mic, we use the theoretical scaling laws derived (for a magnetic reconnection model) by \citet{Maehara2015} and \citet{Namekata2017}:
\begin{equation}\label{scaling}
\tau\propto E^{1/3}B^{-5/3},\qquad
\tau\propto E^{-1/2}L^{5/2},
\end{equation}
where $\tau$ is the flare duration, $E$ is the released flare energy, $B$ is the characteristic magnetic field strength in the flaring region, and $L$ is the length scale of the flaring region; the scaling coefficients were determined from observations of solar flares. For consistency with the results of \citet{Namekata2017}, we characterize a flare with the radiated energy in the optical continuum (assuming $E\simeq E_{\mathrm{cont}}$) and the optical/UV decay timescale (assuming $\tau\simeq\tau_{\mathrm{UV}}$); therefore we consider four flares (F1-F3 and F5) where these parameters were reliably determined. The estimated parameters of the corresponding flaring regions are presented in Table \ref{Flares}. The flaring regions on AT Mic had typical sizes of about $35\,000-70\,000$ km and typical magnetic field strengths of about $80-120$ G (we note that the used scaling laws provide an average magnetic field strength in the flaring volume in the solar/stellar corona, while the field strength at the photospheric level can be much higher).

\begin{figure}
\centerline{\includegraphics{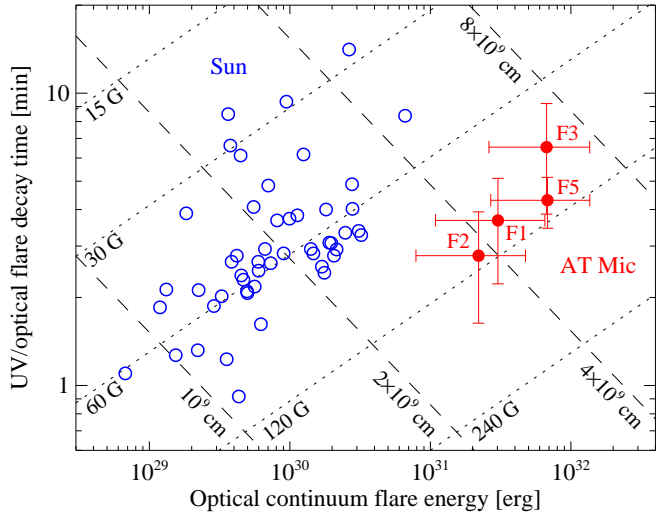}}
\caption{UV/optical flare decay times vs. the optical continuum flare energies for the flares on the Sun \protect\citep[from][shown by empty blue circles]{Namekata2017} and AT Mic (obtained in this work, shown by filled red circles). The dashed and dotted lines represent the theoretical scaling laws by \protect\citet{Namekata2017}. The error bars (shown for the stellar flares only) correspond to the $1\sigma$ level.}
\label{FigETdiagram}
\end{figure}

In Figure \ref{FigETdiagram}, we compare the radiated energies and decay times of the flares detected on AT Mic in this study with the respective parameters of the solar flares (of the GOES classes $\textrm{M1.0}-\textrm{X2.8}$) reported by \citet{Namekata2017}; we also overplot theoretical isolines of characteristic magnetic field strength or flaring region size predicted by the the scaling laws (\ref{scaling}) \citep[cf. similar plots for flares on other stars in the papers of][etc.]{Namekata2017, Namekata2018, Brasseur2019, Tu2020, Tu2021, Ramsay2021}. The flaring region sizes on AT Mic seem to be comparable to or slightly larger than those for the largest solar flares. On the other hand, the magnetic field strengths in the flaring regions on AT Mic are considerably (by a factor of $\sim 1.5-2$) higher than the typical values in the solar flaring regions. A similar conclusion (i.e., that the magnetic fields in stellar flaring regions are typically stronger than in solar ones) was made earlier by \citet{Namekata2017, Namekata2018} for the stellar flares observed with Kepler in the short cadence (1 min) mode and by \citet{Tu2020, Tu2021} for the stellar flares observed with TESS.

The question is open whether the above mentioned differences between the solar and stellar flares are qualitative or quantitative, i.e., whether flares similar to the described here flares on AT Mic can occur on the Sun, albeit rarely. A possible close solar analog of the stellar events was the X9.3 class solar flare on 2017 September 6 that, despite of its high energetics ($\sim 10^{31}$ erg), was relatively short and compact; notably, an anomalously strong magnetic field (up to $\sim 4000$ G in the low corona) was detected in that flare \citep{Anfinogentov2019}. Such rare solar flares with extremely strong magnetic fields deserve further study, because they may bridge the gap between ``ordinary'' solar flares and stellar superflares.\\[30pt]

\section{Summary and conclusions}
We observed the active M-dwarf binary AT Mic with the orbital observatory AstroSat. The target was detected by the SXT telescope in the soft X-ray range ($0.3-7$ keV) and by the UVIT telescope in the far ultraviolet range ($130-180$ nm), with $\sim 10$ ks of simultaneous observations in the two spectral ranges. The main results can be summarized as follows:
\begin{enumerate}
\item
In both the X-ray and UV spectral ranges, we detected quiescent emission and a number of flares. Using the spatially resolved UV observations, we have for the first time identified reliably the components of the binary where the flares occurred: two flares on AT Mic A and four flares on AT Mic B.
\item
The X-ray flares were typically longer than (by a factor of $\sim 2$) and delayed after (by $\sim 5-6$ min) their UV counterparts, demonstrating the Neupert effect.
\item
The X-ray-emitting coronal plasma has been found to be best described by a multi-temperature distribution. In the quiescent state, the emission measure was of $\sim 2.9\times 10^{52}$ $\textrm{cm}^{-3}$ and the average temperature was of $\sim 7$ MK. During flares, both the emission measure and average temperature increased (up to $\sim 4.5\times 10^{52}$ $\textrm{cm}^{-3}$ and $\sim 15$ MK, respectively), corresponding to the plasma temperature in the flaring regions of $\sim 32$ MK.
\item
The abundance of heavy elements in the corona of AT Mic has been found to be much lower than at the Sun ($\sim 0.18$ $Z_{\sun}$ in the quiescent state). During flares, the coronal abundance increased (up to $\sim 0.34$ $Z_{\sun}$), due to chromospheric evaporation.
\item
The detected flares had the radiated energies of $\sim 10^{31}-10^{32}$ erg. The optical continuum emission dominated and was responsible for $\sim 60-65\%$ of the total radiated flare energy.
\item
The estimated sizes of flaring regions on AT Mic ($\sim 35\,000-70\,000$ km) are comparable to or slightly larger than those for the largest solar flares. On the other hand, the estimated magnetic field strengths in the flaring regions on AT Mic ($\sim 80-120$ G) are $\sim 1.5-2$ times higher than those in typical solar flares.
\end{enumerate}

\acknowledgments
This work was supported by the BRICS Multilateral Research and Development Projects-2016 (DST/MRCK/BRICS/PilotCall1/Superflares/2017), the pro\-ject ``Superflares on stars and the Sun'', the Russian Foundation for Basic Research under grant 17-52-80064, and the Ministry of Science and Higher Education of the Russian Federation. 
K.C. is supported by the Research Council of Norway through its Centres of Excellence scheme (project number 262622). 
This research is based on the results obtained from the AstroSat mission of the Indian Space Research Organisation (ISRO), archived at the Indian Space Science Data Centre (ISSDC). 
This work has used the UVIT data processed by the payload operations centre at IIA; the UVIT was built in collaboration between IIA, IUCAA, TIFR, ISRO, and CSA.
This work has used the data from the Soft X-ray Telescope (SXT) developed at TIFR, Mumbai, and the SXT POC at TIFR is thanked for verifying and releasing the data via the ISSDC data archive and providing the necessary software tools.

\appendix
\section{Comparison of different X-ray spectral models}\label{MCMC}
\begin{figure}
\centerline{\includegraphics{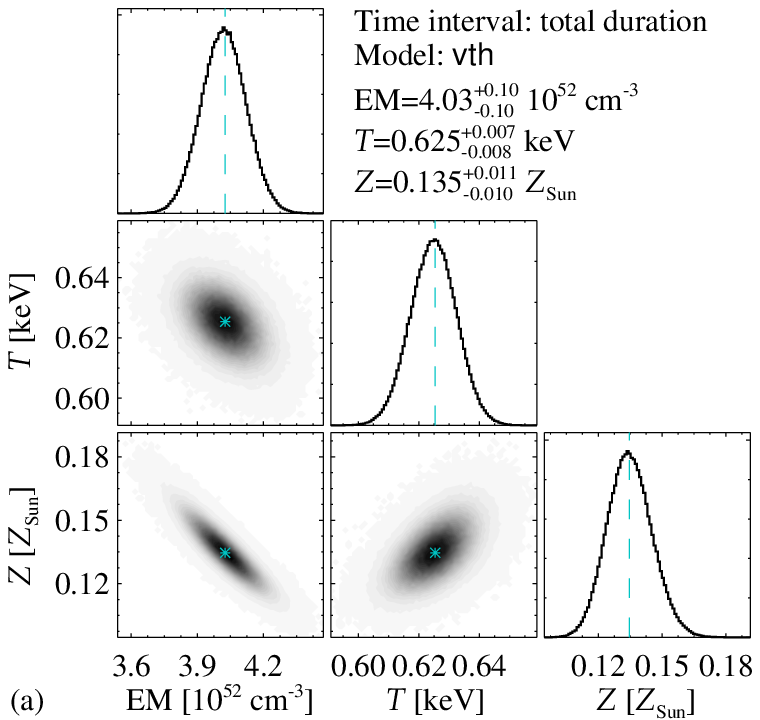}}
\centerline{\includegraphics{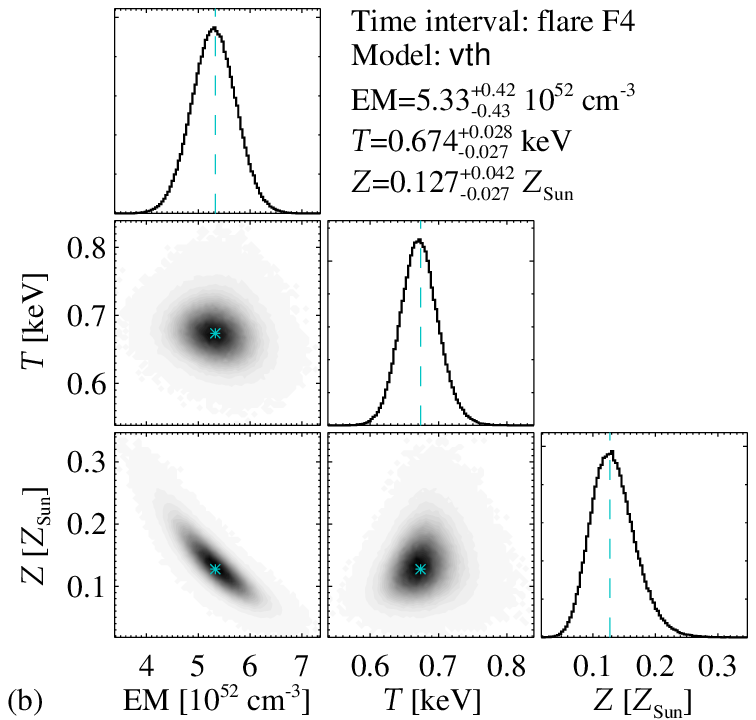}}
\caption{Results of the MCMC analysis of the AstroSat SXT X-ray spectra of AT Mic using the single-temperature optically thin thermal model (\textsf{vth}). The posterior probability distributions are shown, with grey-scale 2D plots demonstrating the joint posterior probability distributions for different combinations of the model parameters (darker areas correspond to higher probability), and histograms demonstrating the marginal posterior probability distributions of the individual parameters. Two time intervals are considered: (a) the total duration of the observations, and (b) flare F4. The most probable values of the model parameters with the $1\sigma$ (68\%) confidence intervals are shown in the panels as well.}
\label{MCMC_vth}
\end{figure}

\begin{figure*}
\centerline{\includegraphics{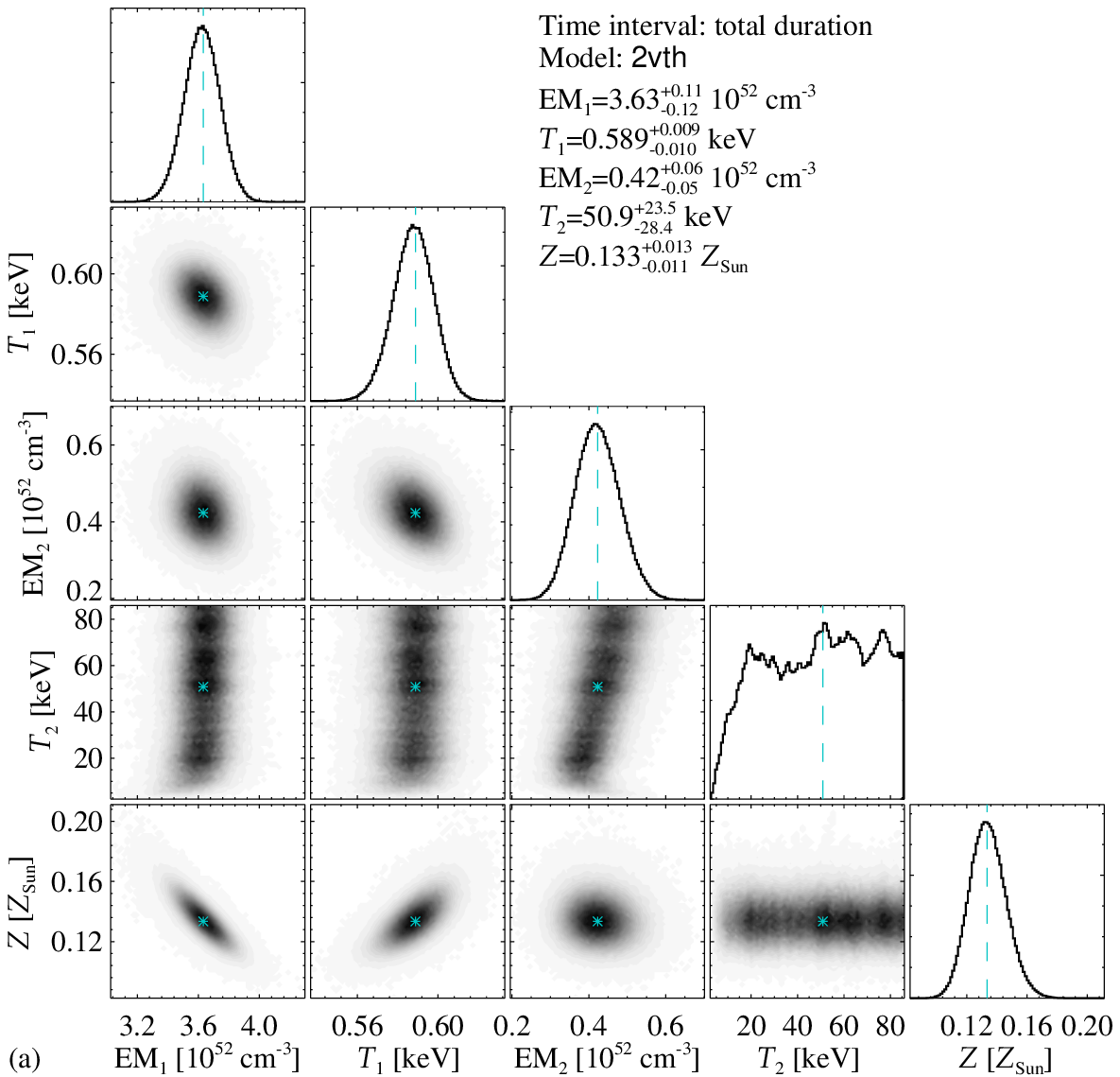}}
\centerline{\includegraphics{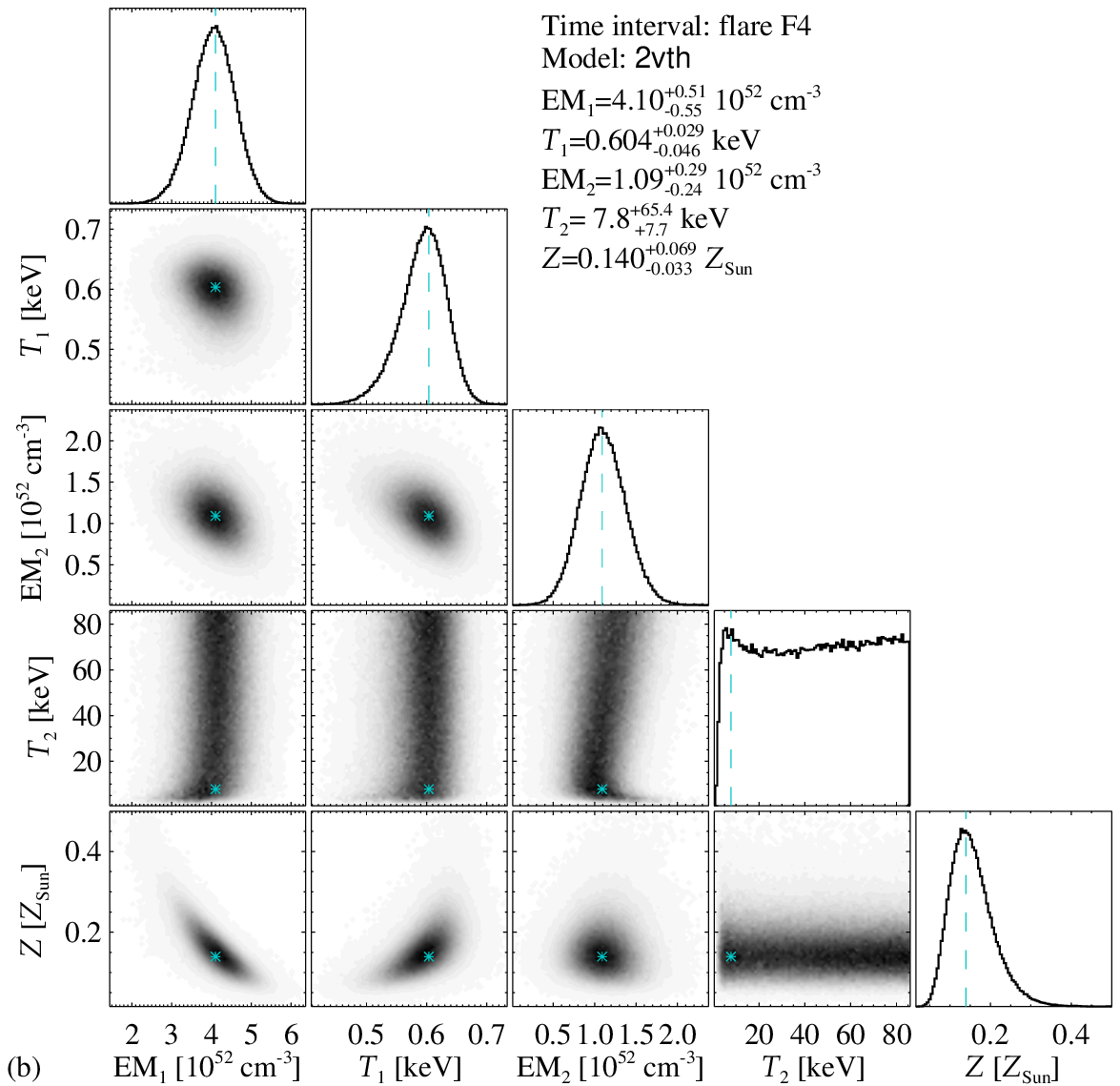}}
\caption{Same as in Figure \protect\ref{MCMC_vth}, for the double-temperature optically thin thermal model (\textsf{2vth}).}
\label{MCMC_2vth}
\end{figure*}

\begin{figure*}
\centerline{\includegraphics{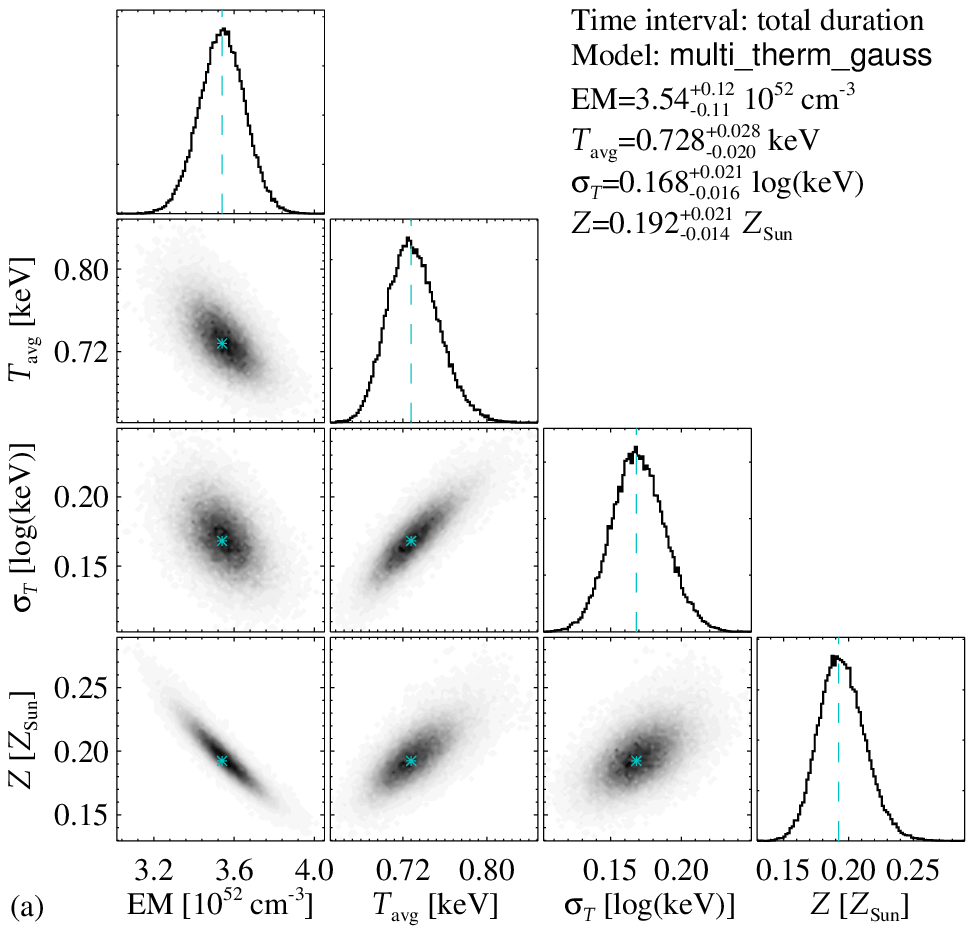}}
\centerline{\includegraphics{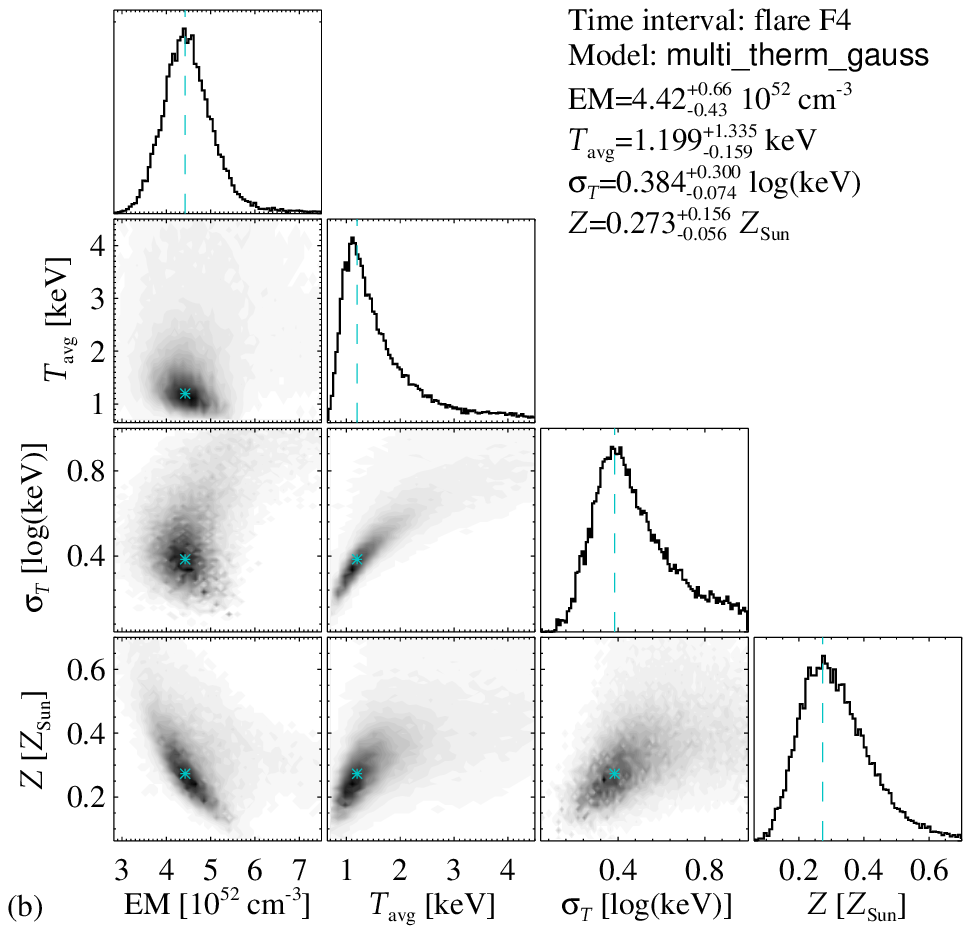}}
\caption{Same as in Figures \protect\ref{MCMC_vth}--\protect\ref{MCMC_2vth}, for the optically thin thermal model with a Gaussian DEM distribution (\textsf{multi\_therm\_gauss}).}
\label{MCMC_gauss}
\end{figure*}

\begin{figure*}
\centerline{\includegraphics{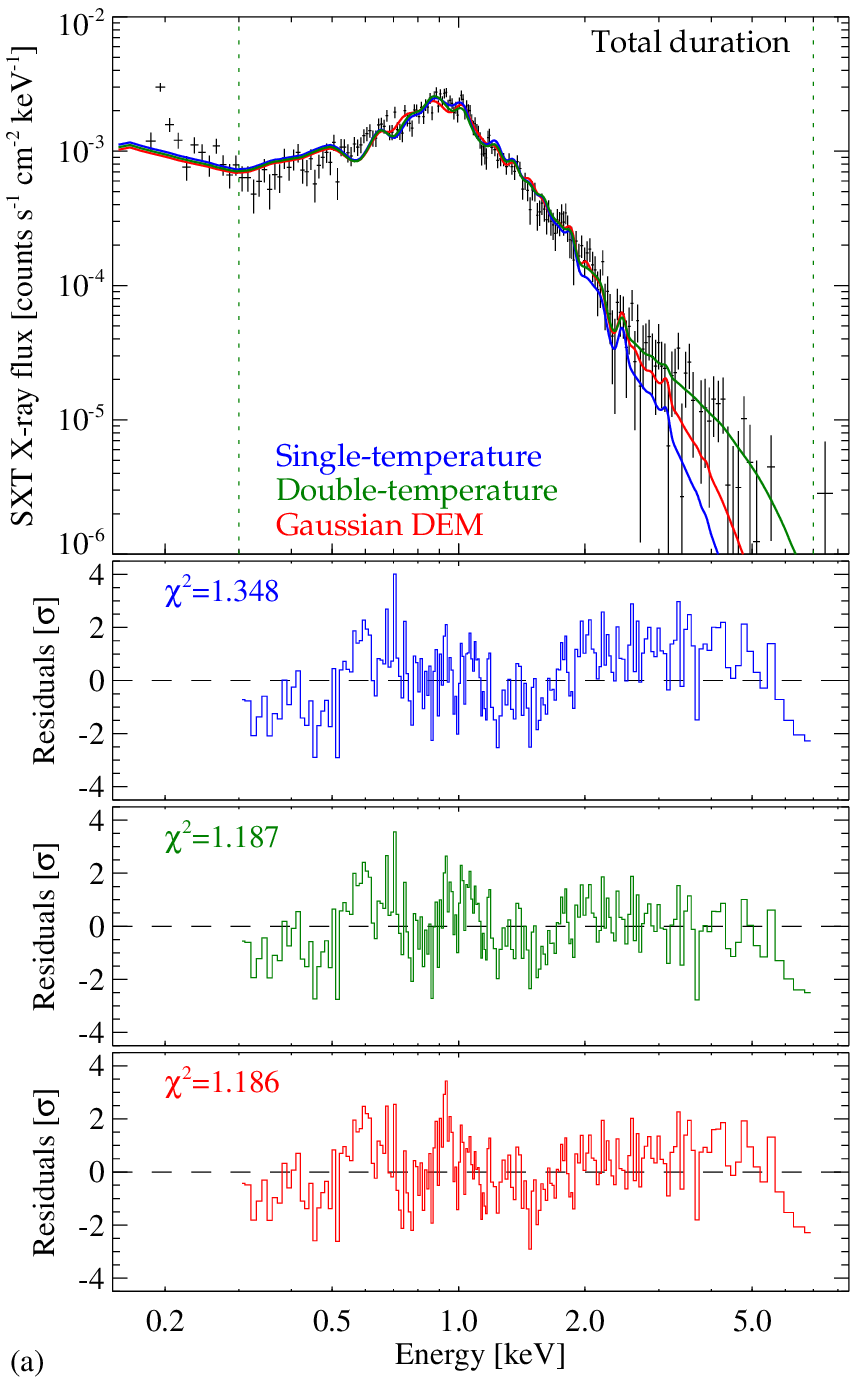}
\includegraphics{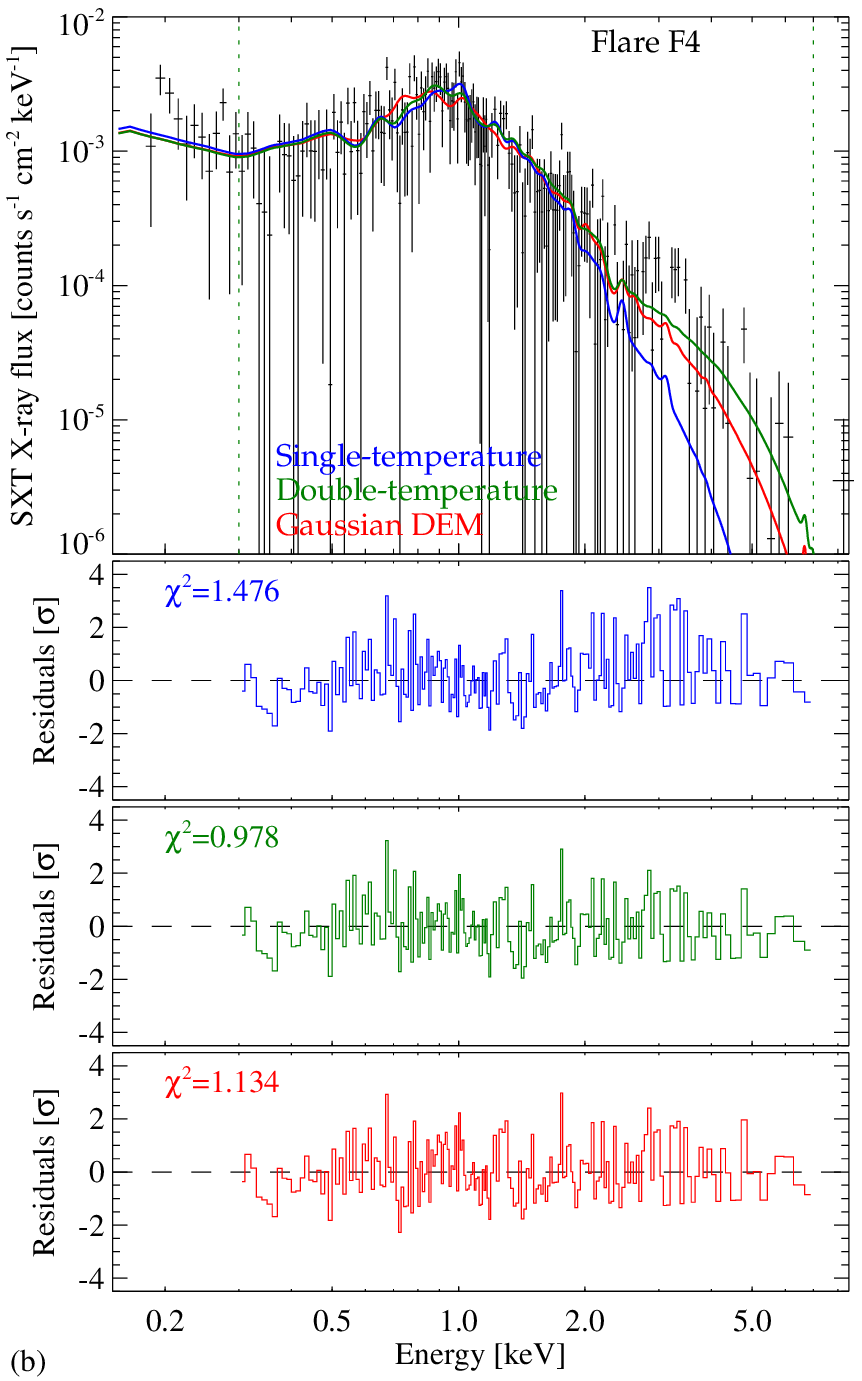}}
\caption{Top panels: X-ray spectra of AT Mic obtained with the AstroSat SXT on 2018 October 3-4, for the entire interval of observations (a) and for the flare F4 (b). At higher energies, the original spectra are rebinned over $2-65$ channels ($0.02-0.65$ keV). The error bars correspond to the $1\sigma$ level. The solid lines represent the best spectral fits with three different plasma emission models obtained using the MCMC technique (see Figures \protect\ref{MCMC_vth}--\protect\ref{MCMC_gauss} for the best-fit model parameters). Bottom panels: normalized residuals of the model spectral fits for the considered plasma emission models, in units of the $1\sigma$ uncertainties; the respective values of the $\chi^2$ statistic are shown as well.}
\label{MCMC_spectra}
\end{figure*}

In order to estimate the parameters of the emitting plasma, we performed fitting of the observed soft X-ray spectra of AT Mic with several spectral models using the Markov Chain Monte Carlo (MCMC) approach; the MCMC sampling implementation by \citet{Anfinogentov2021} was used. In this Section, we present the results obtained for the SXT spectrum averaged over the entire (total) duration of the observations, and for the time interval corresponding to the flare F4 that was the strongest X-ray flare in our observations; the analysis of other time intervals provided qualitatively similar results. We have considered three plasma emission models: a) the single-temperature optically thin thermal model (\textsf{vth}) depending on the emission measure EM, temperature $T$, and metallicity $Z$; b) the double-temperature optically thin thermal model (\textsf{2vth}) depending on the emission measures $\textrm{EM}_1$ and $\textrm{EM}_2$ of two plasma components, the temperatures $T_1$ and $T_2$ of the respective plasma components, and metallicity $Z$; c) the multi-temperature optically thin thermal model with a Gaussian dependence of the differential emission measure (DEM) on the logarithm of the temperature as described by Equations (\ref{DEMmodel}--\ref{TavgDEM}) (\textsf{multi\_therm\_gauss}), depending on the total emission measure EM, the average plasma temperature $\left<T\right>$, the width of the DEM distribution $\sigma_T$, and metallicity $Z$. Figures \ref{MCMC_vth}--\ref{MCMC_gauss} demonstrate the results of the MCMC analysis: the posterior probability distributions for the above mentioned spectral models and time intervals; the best-fit model parameters (i.e., the most probable values of the parameters that also correspond to the minimum values of the $\chi^2$ statistics) are shown as well. Figure \ref{MCMC_spectra} compares the spectral fits obtained using different spectral models.

One can see in Figure \ref{MCMC_spectra} that, in comparison with the single-temperature model, the double-temperature model provides better fits to the observations at high energies ($\gtrsim 2$ keV) and also noticeably lower overall $\chi^2$ values. The secondary plasma component (i.e., the component with the higher temperature) comprises a significant part of the emitting plasma, with the $\textrm{EM}_2/\textrm{EM}_1$ ratio varying from 0.12 for the total duration of the observations up to 0.27 during the flares. These results indicate the presence of at least two plasma components with different temperatures. On the other hand, the temperature of the secondary component $T_2$ in the double-temperature model remains unconstrained, when the values in a very broad range have comparable posterior probabilities (see Figure \ref{MCMC_2vth}) and provide equally good spectral fits; therefore the double-temperature model, while providing useful insights, cannot be used for a reliable quantitative analysis. The uncertainty in determining the secondary component temperature can indicate that this secondary component itself has a complicated composition which cannot be described by a single temperature.

The multi-temperature model with a Gaussian DEM distribution is intermediate (in terms of the number of parameters) between the single-temperature and double-temperature models. It is reduced to a single-temperature model when the width parameter $\sigma_T$ approaches zero, while can account for contributions of multiple plasma components with temperatures in a broad range around the peak temperature for larger values of $\sigma_T$. We have found that the parameters of this model are well constrained by the observations (see Figure \ref{MCMC_gauss}), while the quality of spectral fits (in terms of $\chi^2$) is noticeably better than that for the single-temperature model and comparable to (or only slightly worse than) that for the double-temperature model (see Figure \ref{MCMC_spectra}). Therefore we conclude that, among the considered models, the multi-temperature spectral model \textsf{multi\_therm\_gauss} is the most suitable one to analyze the AstroSat SXT observations of AT Mic.

The presence of multi-thermal plasma in the coronae of red dwarfs, including AT Mic, has been reported in earlier studies as well. E.g., \citet{Robrade2005}, using the XMM-Newton observations, obtained for AT Mic the DEM distribution with a peak at $T_0=7$ MK and FWHM of about 6 MK, which corresponds to $\sigma_T\simeq 0.15$ in Equation (\ref{DEMmodel}); these estimations are consistent with our results for the total duration of the observations (cf. Table \ref{SXTfits}).

\bibliographystyle{aasjournal}
\bibliography{ms2022-0282}
\end{document}